\documentclass[reprint,prb,amsmath,amssymb,superscriptaddress]{revtex4-1}
\usepackage{hyperref}
\usepackage{graphics}
\usepackage{enumerate}

\begin{document}

\title{Quantum Simulation of the Hubbard Model with Dopant Atoms in Silicon}
\author{J. Salfi$^*$}
\affiliation{Centre for Quantum Computation and Communication Technology, School of Physics, The University of New South Wales, Sydney, NSW 2052, Australia.}
\author{J. A. Mol}
\affiliation{Centre for Quantum Computation and Communication Technology, School of Physics, The University of New South Wales, Sydney, NSW 2052, Australia.}
\author{R. Rahman}
\affiliation{Purdue University, West Lafayette, IN 47906, USA.}
\author{G. Klimeck}
\affiliation{Purdue University, West Lafayette, IN 47906, USA.}
\author{M. Y. Simmons}
\affiliation{Centre for Quantum Computation and Communication Technology, School of Physics, The University of New South Wales, Sydney, NSW 2052, Australia.}
\author{L. C. L Hollenberg}
\affiliation{Centre for Quantum Computation and Communication Technology, School of Physics, University of Melbourne, Parkville, VIC 3010, Australia.}
\author{S. Rogge$^\dagger$}
\affiliation{Centre for Quantum Computation and Communication Technology, School of Physics, The University of New South Wales, Sydney, NSW 2052, Australia.}
\date{\today}

\begin{abstract}
In quantum simulation, many-body phenomena are probed in controllable quantum systems. Recently, simulation of Bose-Hubbard Hamiltonians using cold atoms revealed previously hidden local correlations. However, fermionic many-body Hubbard phenomena such as unconventional superconductivity and spin liquids are more difficult to simulate using cold atoms. To date the required single-site measurements and cooling remain problematic, while only ensemble measurements have been achieved. Here we simulate a two-site Hubbard Hamiltonian at low effective temperatures with single-site resolution using subsurface dopants in silicon. We measure quasiparticle tunneling maps of spin-resolved states with atomic resolution, finding interference processes from which the entanglement entropy and Hubbard interactions are quantified. Entanglement, determined by spin and orbital degrees of freedom, increases with increasing covalent bond length.  We find separation-tunable Hubbard interaction strengths that are suitable for simulating strongly correlated phenomena in larger arrays of dopants, establishing dopants as a platform for quantum simulation of the Hubbard model.
\end{abstract}

\maketitle

\section*{Introduction} Quantum simulation offers a means to probe many-body physics that cannot be simulated efficiently by classical computers, using controllable quantum systems to physically realize a desired many-body Hamiltonian\cite{Feynman:1982gn,Cirac:2012jj,Georgescu:2014bg}.  In the analog approach to quantum simulation exemplified by cold atoms in optical lattices\cite{Endres:2011it,Greif:2013kb}, the simulator's Hamiltonian maps to the desired Hamiltonian.  Compared to digital quantum simulation, realized via complex sequences of gate operations\cite{Lanyon:2011jf,Barends:2015jc}, analog quantum simulation is usually carried out with simpler building blocks.  For example, the Heisenberg and Hubbard Hamiltonians of great interest in many-body physics are directly synthesized by cold atoms in optical lattices\cite{Cirac:2012jj,Georgescu:2014bg}.  Although of immense interest and proposed long ago\cite{Stafford:1994gu}, analog simulation of fermionic Hubbard systems has proven to be very challenging\cite{Cirac:2012jj,Georgescu:2014bg}.  The anticipated regime of the intensely debated spin liquid, unconventional superconductivity, and pseudogap\cite{Anderson:1987ii,Balents:2010dsa,Gull:2013hh} has yet to be accessed even for cold atoms.  Here, the required low temperature $T < t/30$ is problematic due to the  weak tunnel-coupling $t$ of cold atoms\cite{Esslinger:2010vf,Greif:2013kb}. Moreover, experimentally resolving individual lattice sites, crucial elsewhere in Bose-Hubbard simulation\cite{Endres:2011it}, remains very challenging in quantum simulation of the Hubbard model.  

Here, we perform atomic resolution measurements resolving spin-spin interactions of individual dopants, realizing an analog quantum simulation of a two-site Hubbard system.  We demonstrate the much desired combination of low effective temperatures, single-site spatial resolution, and non-perturbative interaction strengths of great importance in condensed matter\cite{Anderson:1987ii,Balents:2010dsa,Gull:2013hh}. The dopants' physical Hamiltonian $\mathcal{H}_{\rm sim}$, determined at the time of fabrication\cite{Georgescu:2014bg}, maps to an effective Hubbard Hamiltonian $\mathcal{H}_{\rm sys}=\sum_{i\neq j,\sigma}(t_{ij} c^\dagger_{i\sigma}c_{j\sigma}+\textrm{h.c.})+\sum_{i,\sigma}\mathcal{U}n_{i\uparrow}n_{i\downarrow}$, where $\mathcal{U}$ is the on-site Coulomb repulsion, $c^\dagger_{i\sigma}$ ($c_{i\sigma}$) creates (destroys) a fermion at lattice site $i$ with spin $\sigma$, $n_{i\sigma}=c_{i\sigma}^\dagger c_{i\sigma}$ is the number operator, and h.c. is the Hermitian conjugate. Here, it is desirable to achieve non-perturbative (intermediate) interaction strengths $\mathcal{U}/t$ associated with quantum fluctuations and emergent phenomena\cite{Anderson:1987ii,Balents:2010dsa,Gull:2013hh}, \textit{i.e.}, beyond perturbative Heisenberg interactions (large $\mathcal{U}/t$) realized in photon-based\cite{Ma:2011gv} and ion-based\cite{Friedenauer:2008ii} simulations, and magnetic ions on metal surfaces\cite{Loth:2012dr}. We focus on the system ground state, prepared by relaxation upon cooling\cite{Georgescu:2014bg}, rather than system dynamics.

\begin{figure}
\includegraphics{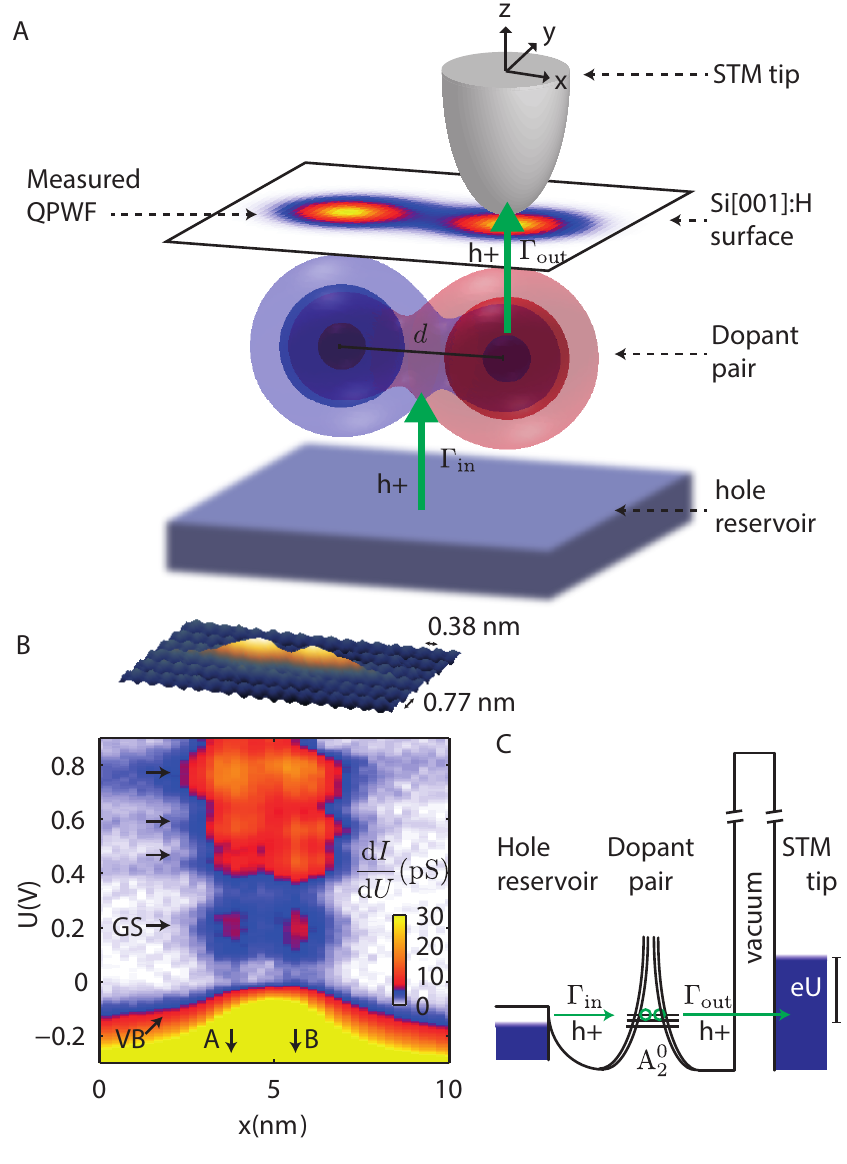}
\caption{\textbf{Spatially resolving coupled-spin states} A. Atomic resolution single hole tunneling probes the interacting states of two coupled acceptor dopants  ($\Gamma_{\rm out}$= tunnel rate to tip, $\Gamma_{\rm out}\ll \Gamma_{\rm in}=$ tunnel rate from reservoir)  The inter-acceptor coupling $t$ obeys $t\gg\hbar\Gamma_{\rm in}$.  ${\rm d}I/{\rm d}U$ measures the interacting states' quasi-particle wavefunction (QPWF), contains interference processes from which we obtain two-body wavefunction amplitudes, and determine the entanglement entropy and effective Hubbard interactions.  B.  Acceptor pair (double-protrusion) in topography at $U=+1.8$ V and $I=300$ pA (top), and spectrally and spatially resolved ${\rm d}I/{\rm d}U$ taken at a bias $U=+2.0$ V where topography is flat apart from atomic corrugation (bottom).  Valence band (VB), 2-hole ground state and 2-hole excited states are indicated.  C.  Effective energy diagram of sequential hole tunneling through 2-hole ground and excited state of coupled acceptors.  }
\label{fig1}
\end{figure}

Because the states of our artificial Hubbard system are coupled and interacting, tunneling spectroscopy locally probes the spectral function.  The spectral function is of key interest in many-body physics because it provides rich information on interactions\cite{Damascelli:2003kq,Fischer:2007fa}, and is highly sought after in future ``cold-atom tunneling microscope'' experiments\cite{Kantian:2015if}.  For our few-body system, the local spectral function describes the quasi-particle wavefunction (QPWF)\cite{Rontani:2005eb,Maruccio:2007kl,Secchi:2012bg,Schulz:2015hv} and the discrete coupled-spin spectrum of the dopants.  We find that interference of atomic orbitals directly contained in the QPWF allows us to quantify the electron-electron correlations and the entanglement entropy.  The entanglement entropy is a fundamental concept for correlated many-body phases\cite{Amico:2008en,Klich:2009jx,Abanin:2012bj,Islam:2015cm} that has thus far evaded measurement for fermions.  In the counterintuitive regime of our experiments, entanglement entropy increases as the valence bond is stretched, as Coulomb interactions overcome quantum tunneling.  In our system, the entanglement entropy is directly related to the Hubbard interactions $\mathcal{U}/t$, and we find that $\mathcal{U}/t$ is tunable with dopant separation, increasing from $4\rightarrow 14$ for $d/a_{\rm B} =2.2\rightarrow 3.7$, where $a_{\rm B}=1.3$ nm is the effective Bohr radius.  This range, of interest to simulate unconventional superconductivity and spin liquids\cite{Anderson:1987ii,Balents:2010dsa,Gull:2013hh}, is realized here due to the large Bohr radii of the hydrogenic states. The semiconductor host allows for electrostatic control of the chemical potential\cite{Fuechsle:2012bl,Koenraad:2011ed}, desirable to dynamically control filling-factor\cite{Anderson:1987ii,Gull:2013hh} but not possible for ions on metal surfaces\cite{Loth:2012dr}.

\section*{Results}
\noindent{\bf Spectroscopy of coupled-spin system} Subsurface boron acceptors in silicon were identified at 4.2 K as individual protrusions\cite{Mol:2013dj,Mol:2015im} (density $\sim10^{11}$ cm$^{-2}$) in constant current images due to resonant tunneling at a sample bias $U=+1.6$ V, and due to the acceptor ion's influence on the valence density of states at $U=-1.5$ V.  The sample was prepared by ultra high vacuum flash annealing at 1200 $^\circ$C and hydrogen termination.  The observed subsurface acceptors had typical depths\cite{Mol:2013dj,Mol:2015im} $< 3$ nm, and correspondingly, a volume density $>25$ times less than the bulk doping, $8\times10^{18}$ cm$^{-3}$.  Pairs of nearby acceptors with $d\lesssim 5$ nm were also found, with a smaller density $\sim10^9$ cm$^{-2}$.  

 The spectrum and spatial tunneling probability of the coupled acceptors were investigated at $T=4.2$ K via single-hole tunneling from a reservoir in the substrate to the dopant pair, to the tip\cite{Mol:2013dj,Mol:2015im} (Fig.~\ref{fig1}A).  For the dopant pair in Fig.~\ref{fig1}B (top), ${\rm d}I/{\rm d}U$ measured along the inter-dopant axis (Fig.~\ref{fig1}B, bottom) contains a peaks for each state entering the bias window, at $U\approx 0.2$ V, 0.45 V, 0.55 V and 0.8 V.  Consistent with our single-acceptor\cite{Mol:2013dj} and single-donor\cite{Salfi:2014kaa} measurements near flat-band bias conditions, the bias for each peak in the spectrum (Fig.~\ref{fig1}B, bottom) is independent of tip position.  This rules out distortion of our quantum state images by inhomogenous tip-induced potentials\cite{Teichmann:2008bh} observed in other multi-dopant systems\cite{Teichmann:2011gs}. These results can be attributed to weak electrostatic control by the tip (Fig.~\ref{fig1}C) and the states' proximity to flat-band\cite{Mol:2013dj,Mol:2015im,Salfi:2014kaa}, though a large tip radius may also play a role.  

The spectral and spatially resolved measurements (Fig.~\ref{fig1}B) directly demonstrate that the holes are interacting, as follows.  First, two peaks centred on dopant ions A or B are resolved in real space (Fig.~\ref{fig1}B).  Second, energy differences between the peaks resolved in real space are smaller than the $\sim350$ $\mu$eV thermal resolution.  However, for orbitals at the same energy to not interact, their overlap must vanish.  Since the measured orbitals have a strong overlap, the sites are tunnel coupled, irrespective of the details of the tunneling current profile.  The number of states observed, their energy differences, and their energies relative to the Fermi energy confirm that they observed states are two-hole states (Fig.~\ref{fdIdUfit} and ~\ref{fSplitting}).  \\

\begin{figure}
\includegraphics{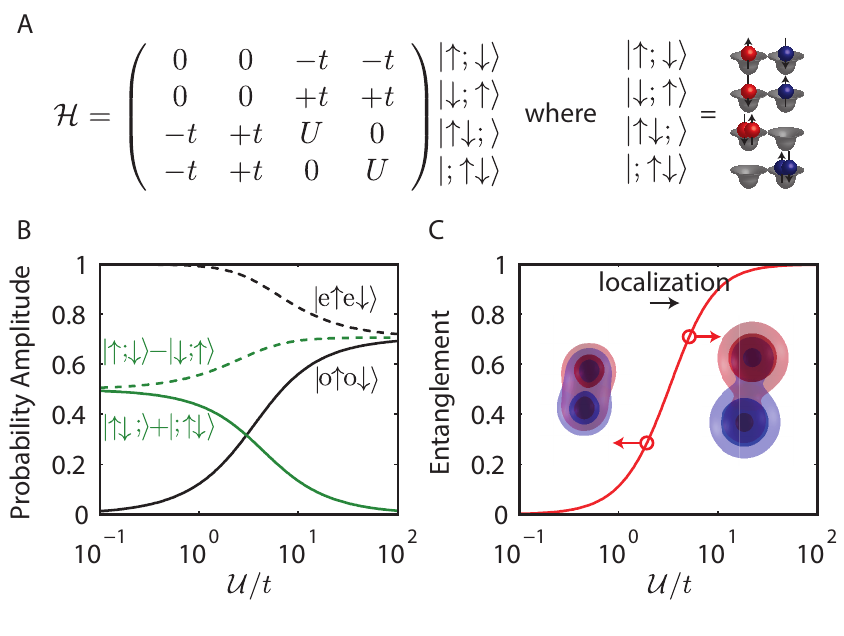}
\caption{\textbf{Hubbard Interactions and Entanglement Entropy} A. Two-site Hubbard Hamiltonian in the subspace of the ground state, with tunnel coupling $t$ hybridizing singly and doubly occupied configurations, for sites $A$ (red orbital) and $B$ (blue orbital).  B. Dependence of probability amplitudes on interactions $\mathcal{U}/t$: $\gamma_{\rm c}$ (green dashed) and $\gamma_{\rm i}$ (green solid) for configurations $(\left|\uparrow;\downarrow\right\rangle-\left|\downarrow;\uparrow\right\rangle)$ and $(\left|\uparrow\downarrow;\right\rangle+\left|;\uparrow\downarrow\right\rangle$), and $\gamma_{\rm ee}$ and $\gamma_{\rm oo}$ for $\left|{\rm e}_\uparrow {\rm e}_\downarrow \right\rangle$ and $\left|{\rm o}_\uparrow {\rm o}_\downarrow \right\rangle$ respectively.  C. Entanglement entropy $\mathcal{S}$ increases with increasing Hubbard interactions $\mathcal{U}/t$.  This occurs because of localization of red and blue orbitals associated with spins in the singlet, as illustrated in the insets. }
\label{fig2}
\end{figure}

\noindent{\bf Correlations, Entanglement, and Hubbard Interactions} The ground state of a Hubbard model with non-perturbative interactions is governed by $\mathcal{H}$ in Fig.~\ref{fig2}A in the subspace of $\left|\uparrow;\downarrow\right\rangle=c^\dagger_{A\uparrow}c^\dagger_{B\downarrow}\left|0\right\rangle$, $\left|\downarrow;\uparrow\right\rangle=c^\dagger_{A\downarrow}c^\dagger_{B\uparrow}\left|0\right\rangle$, $\left|\uparrow\downarrow;\right\rangle=c^\dagger_{A\uparrow}c^\dagger_{A\downarrow}\left|0\right\rangle$ and $\left|;\uparrow\downarrow\right\rangle=c^\dagger_{B\uparrow}c^\dagger_{B\downarrow}\left|0\right\rangle$, where $c^\dagger_{i\sigma}$ creates a localized electron on site $i\in\{A,B\}$ with spin $\sigma\in\{\uparrow,\downarrow\}$, and $\left|0\right\rangle$ is the vacuum state.  The ground state is a superposition $\left|\Psi_{\rm S}\right\rangle=\gamma_{\rm c}(\left|\uparrow;\downarrow\right\rangle-\left|\downarrow;\uparrow\right\rangle)+\gamma_{\rm i}(\left|\uparrow\downarrow;\right\rangle+\left|;\uparrow\downarrow\right\rangle)$, where $\gamma_{\rm c}$ ($\gamma_{\rm i}$) is the probability amplitude for a covalent (ionic) configuration (Fig.~\ref{fig2}B).  Rewriting the state in a basis of even and odd orbitals, $\left|\Psi_{\rm S}\right\rangle=\gamma_{\rm ee}\left|{\rm e}_\uparrow {\rm e}_\downarrow \right\rangle-\gamma_{\rm oo}\left|{\rm o}_\uparrow {\rm o}_\downarrow \right\rangle$, where $\gamma_{\rm ee}$ ($\gamma_{\rm oo}$) is the probability amplitude of the ``even/even'' (``odd/odd'') configuration. 

In limit of small tunnel couplings ($\mathcal{U}/t\rightarrow\infty$, Fig.~\ref{fig2}B) the Hubbard system may be described by perturbative Heisenberg spin interactions. For vanishing $\mathcal{U}/t$, the ground state is a Heitler-London singlet of localized spins, $\left|\Psi_{\rm S}\right\rangle=2^{-1/2}(\left|\uparrow;\downarrow\right\rangle-\left|\downarrow;\uparrow\right\rangle)$, with no contributions from $\left|\uparrow\downarrow;\right\rangle$ and $\left|;\uparrow\downarrow\right\rangle$.  Due to vanishing wavefunction overlap the electrons can be associated with sites A and B (they are distinguishable\cite{Schliemann:2001ea,Ghirardi:2004dl,Amico:2008en}), and the spin at site A depends on the spin at site B as for a maximally entangled Bell state.  In the limit of vanishing interactions ($\mathcal{U}/t\rightarrow 0$, Fig.~\ref{fig2}B) corresponding to a tight-binding approximation, the spins delocalize and $\left|\Psi_{\rm S}\right\rangle=\tfrac{1}{2}(\left|\uparrow;\downarrow\right\rangle-\left|\downarrow;\uparrow\right\rangle)+\tfrac{1}{2}(\left|\uparrow\downarrow;\right\rangle+\left|;\uparrow\downarrow\right\rangle)$.  In a molecular orbital (MO) basis, the ground state $\left|\Psi_{\rm S}\right\rangle=\left|{\rm e}_\uparrow{\rm e}_\downarrow\right\rangle$, which is a single Slater determinant.  Although this state is a singlet (one spin up, one spin down) due to fundamental indistinguishability, the electrons can be ascribed independent properties because they occupy the same orbital, and the state is uncorrelated\cite{Schliemann:2001ea,Ghirardi:2004dl,Amico:2008en}. 

In the regime of intermediate $\mathcal{U}/t$ where tunneling and Coulomb interactions compete non-perturbatively\cite{Anderson:1987ii,Gull:2013hh,Cirac:2012jj,Georgescu:2014bg}, tunnelling hybridizes the doubly-occupied configurations $\left|\uparrow\downarrow;\right\rangle$ and $\left|;\uparrow\downarrow\right\rangle$ into the ground state, such that the particles lose their individual identities.  Here, the von Neumann entanglement entropy quantifies genuine entanglement (inter-dependency of properties), distinguishing it from exchange-correlations due to indistinguishability\cite{Ghirardi:2004dl,Amico:2008en,Islam:2015cm}.  Employing the convention\cite{He:2005in} $\mathcal{S}=0$ (1) for zero (maximal) entanglement, $\mathcal{S}=-|\gamma_{\rm ee}|^2\log_2|\gamma_{\rm ee}|^2-|\gamma_{\rm oo}|^2\log_2|\gamma_{\rm oo}|^2$ increases as $\mathcal{U}/t$ increases and coherent localization occurs (Fig.~\ref{fig2}C), saturating at value of 1.  

\begin{figure*}
\includegraphics{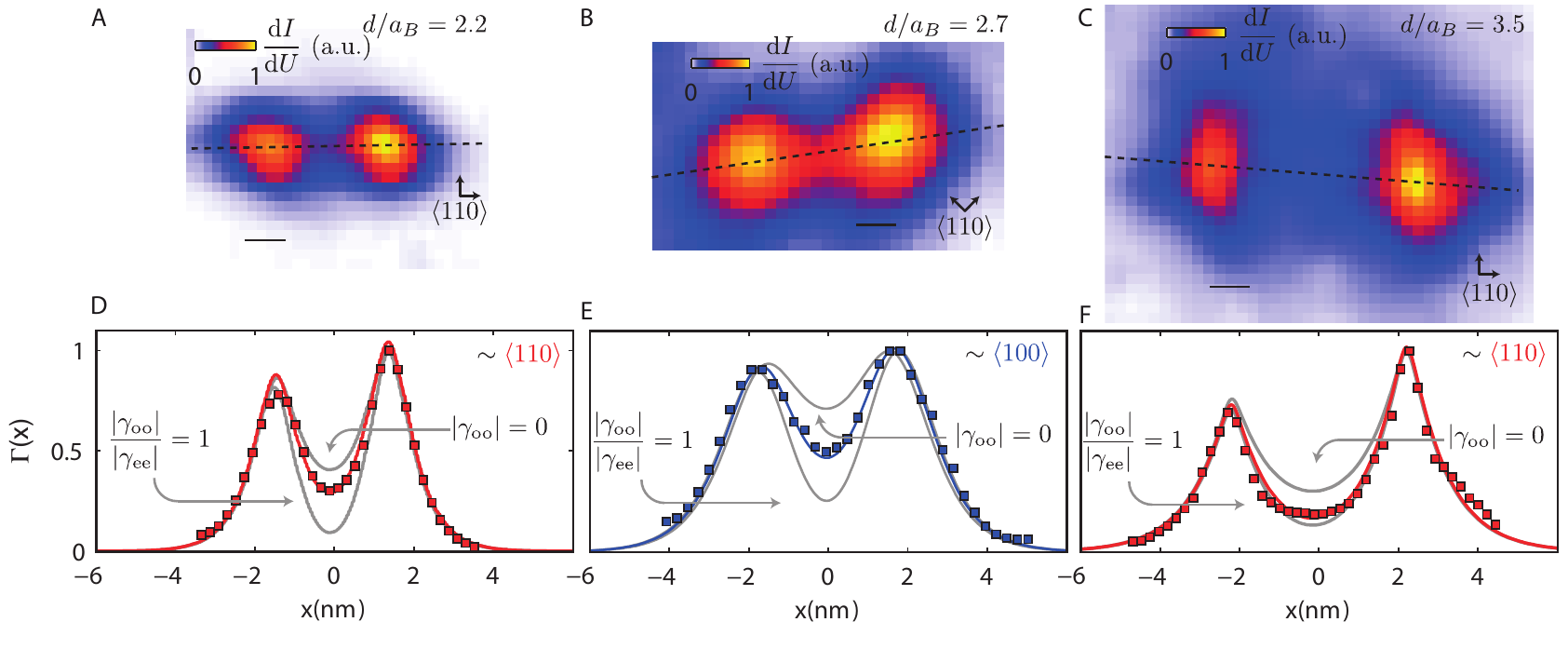}
\caption{\textbf{Resolving Interference Processes in Quasi-Particle Wave Function} A. Experimentally measured, normalized tunneling probability $\Gamma\propto{\rm d}I/{\rm d}U$ to tip, for $d=2.2a_{\rm B}$ ground state.  Arrows denote $110$ crystal directions.  B. Same as (A), for $d=2.7a_{\rm B}$. C. Same as (A), for $d=3.5a_{\rm B}$.  D. Normalized experimental line profile (coloured squares) of $\Gamma(x)$ for $d=2.2a_{\rm B}$ and least-squares fit (coloured line) to QPWF correlated singlet model.  Lower and upper grey lines are line profiles of maximally and minimally correlated states, obtained from least square fits.  The maximally correlated state deviates from the mean of $|\phi_{e}(\mathbf{r})|^2$ and $|\phi_{o}(\mathbf{r})|^2$ because of the different normalization coefficients of even and odd linear combinations.  E.  Same as (D), for $d=2.7a_{\rm B}$.  F.  Same as (D), for $d=3.5a_{\rm B}$.  Scale: 1 nm.  }
\label{fig3}
\end{figure*}

\noindent{\bf }We now discuss the spatial tunneling maps of the two-hole ground states for different inter-acceptor distances.  Obtained by integrating the lowest voltage ${\rm d}I/{\rm d}U$ peak, the maps are shown in Fig.~\ref{fig3}A, \ref{fig3}B, and \ref{fig3}C for distances $d/a_{\rm B}=2.2,2.7$ and $3.5$ ($a_{\rm B}=1.3$ nm) having orientations $\pm 2^\circ$ from $\left\langle 110 \right\rangle$, $8\pm 2^\circ$ from $\left\langle 100 \right\rangle$ and $3\pm 2^\circ$ from $\left\langle 110 \right\rangle$, respectively.  The multi-nm spatial extent of the states reflects the extended wave-like nature of the acceptor-bound holes, owing to their shallow energy levels, which contrasts Mn ions on GaAs surfaces\cite{Kitchen:2006hv}, magnetic ions on metals\cite{Loth:2012dr}, and Si(001):H dangling bonds\cite{Schofield:2013kt}. Consequently, their envelopes are amenable to effective-mass analysis with lattice frequencies filtered out\cite{Yakunin:2004et,Rontani:2005eb,Maruccio:2007kl,Koenraad:2011ed}.  Consistent with measurements of single acceptors at similar depths on resonance at flatband\cite{Mol:2013dj,Mol:2015im}, the states have predominantly s-like envelopes with slight extension along [110] directions, as expected when symmetry is not strongly perturbed by the surface. Depths of the $d/a_{\rm B}=2.7$ and $d/a_{\rm B}=3.5$ pairs were estimated to be $\sim 0.9$ nm, and for $d/a_{\rm B}=2.2$, $\sim 0.6$ nm (see Fig.~\ref{fDepth}). 

We employed full-configuration interaction calculations of the singlet ground-state $\left|\Psi_{\rm S}\right\rangle$ to confirm that Coulomb correlations of coupled acceptors influence the ground state in a way that mimics the $S=1/2$ Hubbard model.  In particular, for $d/a_{\rm B} \sim 2$, $\left|\Psi_{\rm S}\right\rangle$ is predominantly composed of $c^\dagger_{{\rm e},3/2}c^\dagger_{{\rm e},-3/2}|0\rangle$, a singlet of two even $\pm$``3/2'' spin MOs.  With increasing $d$, interactions enhance the probability amplitude of the $c^\dagger_{{\rm o},3/2}c^\dagger_{{\rm o},-3/2}|0\rangle$ singlet with two odd orbitals, analogous to the Hubbard Hamiltonian (Fig.~\ref{fig2}B).  The spins $\pm$``3/2'' are predominantly composed of $\left|3/2,\pm 3/2\right\rangle$ valence band Bloch states.  In particular, the low-lying $\pm$``1/2'' spin excitations of each acceptor\cite{Mol:2015im}, which are predominantly composed of $\left|3/2,\pm1/2\right\rangle$ Bloch states, do not qualitatively change the description.  We also note that for $d/a_{\rm B}\gtrsim2$, the MOs are essentially linear combinations atomic orbitals having the effective Bohr radii of single acceptors.

Single-hole tunneling transport through our coupled dopant system locally probes the spectral quasi-particle wavefunction\cite{Rontani:2005eb,Maruccio:2007kl,Secchi:2012bg}. When $\Gamma_{\rm out}\ll\Gamma_{\rm in}$ (Fig.~\ref{fig1}A), the single-hole tunneling rate is essentially governed by $\Gamma_{\rm out}$, the tunnel-out rate\cite{Salfi:2014kaa}. In the present case, single-hole tunneling from the two-hole system to a single-hole final state $|f\rangle=c_f^\dagger|0\rangle$ (Fig.~\ref{fig1}) contributes $\Gamma^f_{\rm out}(\mathbf{r})=|\langle f|\hat{\Psi}(\mathbf{r})|\Psi_{\rm S}\rangle|^2$, where $\langle f|\hat{\Psi}(\mathbf{r})|\Psi_{\rm S}\rangle$ is the QPWF, $\hat{\Psi}(\mathbf{r})=\sum_j\phi_j(\mathbf{r})c_j$ is the field operator, $c^\dagger_j$ creates a single-hole MO eigenstate $\phi_j(\mathbf{r})$ of the system\cite{Rontani:2005eb}, and the total tunnel rate is $\Gamma(\mathbf{r})=\sum_f \Gamma_{\rm out}^f(\mathbf{r})$.

From the QPWF description of coupled dopants, we obtain a spatial tunneling probability $\Gamma(\mathbf{r},|\gamma_{\rm ee}|,|\gamma_{\rm oo}|)\propto|\gamma_{\rm ee}|^2|\phi_{\rm e}(\textbf{r})|^2 + |\gamma_{\rm oo}|^2|\phi_{\rm o}(\textbf{r})|^2$ for the ground state.  Here, $|\gamma_{\rm ee}|^2$ and ($|\gamma_{\rm oo}|^2$) contain constructive (destructive) interference corresponding to even (odd) linear combinations of atomic orbitals $\phi_{\rm e}(\mathbf{r}_1)$ ($\phi_{\rm o}(\mathbf{r}_1)$) (note: $|\gamma_{\rm ee}|^2+|\gamma_{\rm oo}|^2=1$).  To obtain $|\gamma_{\rm oo}|^2$, data were fit to $\Gamma(\mathbf{r},|\gamma_{\rm ee}|,|\gamma_{\rm oo}|)$, assuming linear combinations of parametrized s-like atomic orbitals for $\phi_{\rm e}(\mathbf{r})$ and $\phi_{\rm o}(\mathbf{r})$ appropriate for subsurface acceptors. The QPWF and atomic orbitals are described in Figs.~\ref{fTheoryQPWF}, \ref{fSingleBoron} and \ref{fDoubleBoron}. 

The least-squares fits in Figs.~\ref{fig3}D, \ref{fig3}E, \ref{fig3}F (colored lines) of $\Gamma(\mathbf{r},|\gamma_{\rm ee}|,|\gamma_{\rm oo}|)$ are in good agreement with data (squares), for $d/a_{\rm B}=2.2$, $2.7$ and $3.5$.  For comparison with the data, grey curves are shown for both the uncorrelated (maximally correlated) state with $|\gamma_{\rm oo}|=0$ ($|\gamma_{\rm oo}|/|\gamma_{\rm ee}|=1$) in Fig.~\ref{fig3}D, \ref{fig3}E, and \ref{fig3}F.  We note that all three separations exhibit interaction effects at the midpoint of the ions, where the quantum interference is strongest.  We obtain $|\gamma_{\rm oo}|^2=0.12\pm0.06 $, $0.23\pm0.07$, and $0.39\pm0.08$ for $d/a_{\rm B}=2.2,2.7$ and $3.5$.  Data taken at higher tip heights gave identical results to within experimental errors (see Fig.~\ref{fCorr} and \ref{fIVZ}), independently verifying that the tip does not influence our results.  

\begin{figure}
\includegraphics{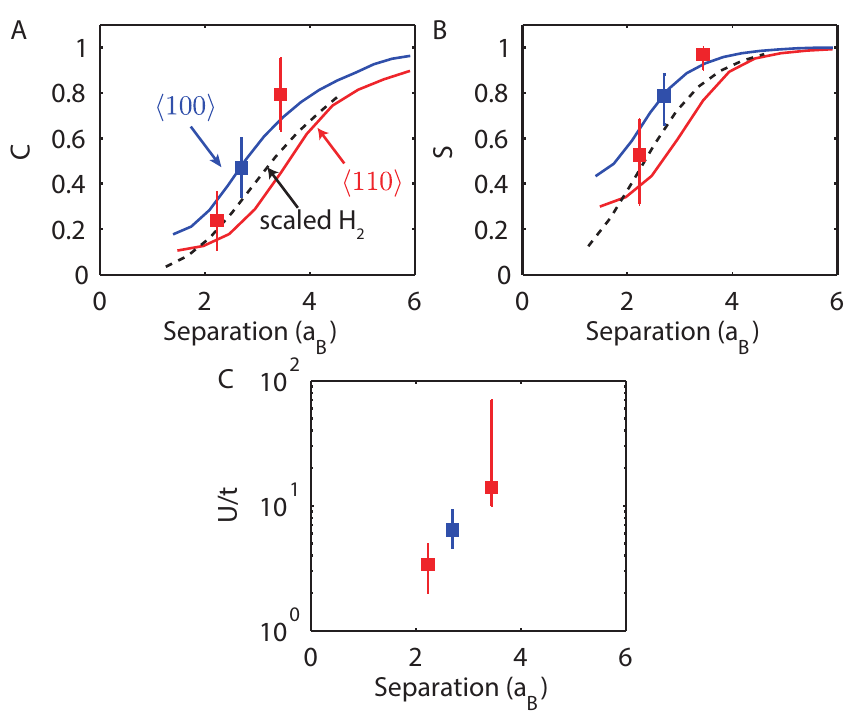}
\caption{\textbf{Entanglement Entropy and Hubbard interactions} A. Quantum correlations $C$ vs. $d$.  Theory predictions are shown for coupled acceptors with $\left\langle 110 \right\rangle$ orientations (red line) and $\left\langle 100 \right\rangle$ (blue line), alongside scaled H$_2$ (dashed black line).  Predicted localization is suppressed (enhanced) along  $\left\langle 110 \right\rangle$ ($\left\langle 100 \right\rangle$) relative to molecular hydrogen (H$_2$), due to valence band anisotropy, which enhances (suppresses) $t$.  B.  Same as (A), for the entanglement entropy $\mathcal{S}$.  C.  Experimentally estimated Hubbard interactions. Error bars denote 95 \% confidence intervals. }
\label{fig4}
\end{figure}

The Coulomb correlations, embodied both in $\mathcal{C}=2|\gamma_{\rm oo}|^2$ (Fig.~\ref{fig4}A) and the entanglement entropy $\mathcal{S}=-|\gamma_{\rm ee}|^2\log_2|\gamma_{\rm ee}|^2-|\gamma_{\rm oo}|^2\log_2|\gamma_{\rm oo}|^2$ (Fig.~\ref{fig4}B), could be evaluated directly from the fit, and both increase with increasing $d$.  The one-to-one mapping from $\mathcal{S}$ to $\mathcal{U}/t$ (Fig.~\ref{fig2}C) was used to determine the effective Hubbard interactions from the entanglement entropy in Fig.~\ref{fig4}B. We obtain $\mathcal{U}/t\approx3.5$, 6.4, and 14, for $d/a_{\rm B}=2.2$, $2.7$, and $3.5$ respectively (Fig.~\ref{fig4}C), which increase as the tunnel coupling decreases.  

We conclude the analysis of the QPWFs with some critical remarks on correlations extracted from our fitting model, recalling that the large spatial overlap of the spectrally overlapping acceptor-bound holes directly shows their states are tunnel coupled.  First, the Coulomb correlations have a systematic effect on interference in the QPWF such that the least-squares error is significantly worse if $|\gamma_{\rm oo}|^2$ is forced to zero in the fitting model (Table~\ref{tSSE}).  Second, if applied to very far apart dopants where the ground state can still be resolved, our fitting model would not give a spurious result that the two dopants are highly correlated.  This follows because the difference between $|\phi_{\rm e}(\mathbf{r})|^2$ and $|\phi_{\rm o}(\mathbf{r})|^2$, which reflects the interference of atomic orbitals and is used to detect correlations, tends to zero as $d/a_{\rm B}$ increases.  Data (Fig.~\ref{fig3}A-C) presented here are for coupled dopants that we found to be (i) well isolated from other dopants or dangling bonds, and (ii) at identical depths, as evidenced by the spatial extent and brightness of the atomic orbitals.  When the latter is not satisfied, the atomic levels can be detuned, introducing more parameters to the fit. \\

{\bf Comparison with theory} These experimental results obey the trends predicted by our theory calculations for the spin-orbit coupled valence band.  Predictions in Fig.~\ref{fig4}A and \ref{fig4}B for displacements along $\left\langle 100 \right\rangle$ (blue solid line) and $\left\langle 110 \right\rangle$ (red solid line) both show increasing correlations and entanglement with increasing dopant separation.  Moreover, we find that the observed and predicted entanglement entropy qualitatively reproduce a single-band model (Fig.~\ref{fig4}A, \ref{fig4}B, dashed lines).  This result implies that inter-hole Hubbard interactions follow an essentially hydrogenic trend with atomic separation, even for non-perturbative interactions $\mathcal{U}/t=4\rightarrow14$. 
  
The hydrogenic nature of $\mathcal{S}$ and $\mathcal{U}/t$ persists in spite of the $\pm$``1/2'' spin excited states of a single acceptors. Such $\pm$``1/2'' single-acceptor excited states states are found nominally $\Delta\sim 1-2$ meV above the $\pm$``3/2'' spin ground state due to inversion symmetry breaking at the interface\cite{Mol:2015im}. Although $t>\Delta$, $\mathcal{S}$ and $\mathcal{U}/t$ remain hydrogenic in our calculations because the ``1/2'' spin excited state has an s-like envelope whose spatial extent is similar to (1) the s-like $\pm$``3/2'' ground state and (2) the scaled hydrogenic ground state.  Otherwise, single particle $\pm$``1/2'' states would hybridize stronger than single particle $\pm$``3/2'' states, form the 2-hole singlet at smaller separations, and localize more slowly relative to molecular hydrogen with increasing $d$.  Furthermore, the polarization of the $\pm$``3/2'' and $\pm$``1/2'' states into $\left|3/2,\pm3/2\right\rangle$ and $\left|3/2,\pm1/2\right\rangle$ components, respectively, limits the mixing of $\pm$``1/2'' states into the ground state.  \\

\begin{figure}
\includegraphics{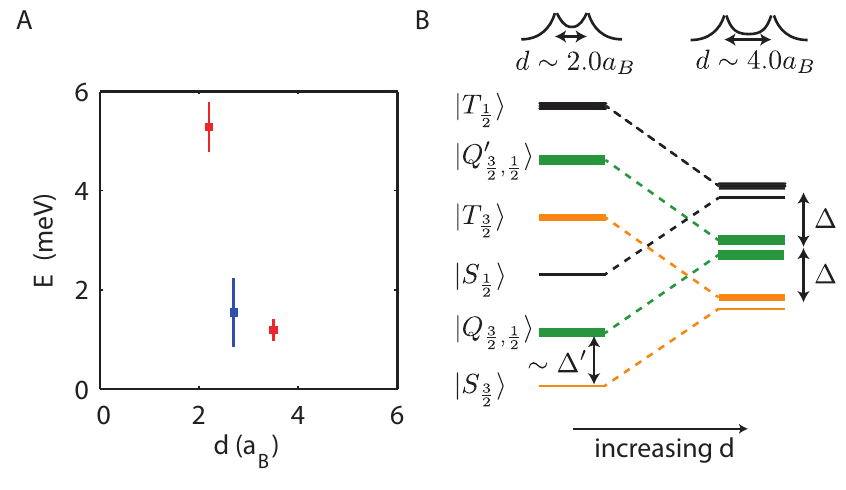}
\caption{\textbf{Coupled-spin excitation spectrum} A. Measured energy of first excited state relative to ground state.  B. Schematic level diagram of coupled acceptors, reflecting theory calculations, as a function of inter-acceptor distance $d/a_{\rm B}$.  Singlets $\left|S_{m_J}\right\rangle$ and triplets $\left|T_{m_J}\right\rangle$ are present for interactions between two holes of $m_J=\pm$``3/2'' spin (orange) and two holes of $m_J=\pm$``1/2'' (black) spin.  States $|Q_{3/2,1/2}\rangle$ and $|Q'_{3/2,1/2}\rangle$ are sets of four closely spaced levels (green) with one ``3/2`` spin hole, and one ``1/2'' spin hole. Error bars denote 95 \% confidence intervals. }
\label{fig5}
\end{figure}

{\bf Spin excited states and effective temperature} Finally, we discuss the observed excited states, which confirm that the inter-acceptor tunnel-coupling dominates thermal and tunnel-coupling effects of the reservoir.  The energies of the states were determined by fitting the single-hole transport lineshapes\cite{Voisin:2015gl} of the coupled acceptors (Fig.~\ref{fdIdUfit} and \ref{fSplitting}).  For the first excited state we found $5.2\pm0.6$ meV and $1.2\pm0.2$ meV for $d/a_{\rm B}=2.2$ and $3.5$ respectively ($\sim\langle110\rangle$ orientation), and $1.6\pm 0.7$ meV for $d/a_{\rm B}=2.7$ ($\sim\langle100\rangle$ orientation).  Shown in Fig.~\ref{fig5}A, these energies are too small to add another hole, which would require $\approx 50$ meV for an acceptor in bulk silicon.  However, the energies agree well with our predictions for two-hole excited states of coupled hole spins $\pm\textrm{``3/2''}$ and $\pm\textrm{``1/2''}$, \textit{i.e.}, $8.5$ meV and $1.5$ meV for $d=2.2a_{\rm B}$ and $d=3.5a_{\rm B}$ ($\langle110\rangle$ orientation), and $2.0$ meV ($\langle100\rangle$ orientation).  Here we note that some of the predicted coupled-spin excited states (Fig.~\ref{fig5}B) are unconventional: a singlet $\left| S_{m_J}\right\rangle$ and triplet $|T_{m_J}\rangle$ of two ``3/2'' holes (orange lines) and two ``1/2'' holes (black lines), where $\left|S_{3/2}\right\rangle$ is the ground state for all separations.  More subtly, two manifolds $|Q^i_{3/2,1/2}\rangle$, $|Q'^i_{3/2,1/2}\rangle$, $i=1\dots4$, containing four states are predicted (green lines), where one $\pm$``3/2'' spin level and one $\pm$``1/2'' spin level is occupied.  For $d/a_{\rm B}=2.2$ and $2.7$ ($d/a_{\rm B}=3.5$), the measured energies are in better agreement with predictions for $|Q^i_{3/2,1/2}\rangle$ ($|T_{3/2}\rangle$) excitations.  

The inter-acceptor tunnel couplings $t$ (ratios $t/T$) were estimated to be 12 meV (30), 7 meV (20), and 3.5 meV (10) for $d/a_{\rm B}=2.2,2.7$ and $3.5$ respectively, at $T=4.2$ K.  Such couplings $t$ exceed the reservoir coupling $\Gamma_{\rm in}$ (Table~\ref{tCouplingTable}) to the substrate by more than 50X.  Combined with bias $U\sim0.2-0.3$ V needed to bring the level into resonance, this rules out coherent interactions with substrate and tip reservoirs\cite{Neel:2011ch}.  Note that the measured energy splittings imply small thermal excited-state populations of $\lesssim 10^{-5}$, $\lesssim 10^{-2}$, and $\lesssim 10^{-1}$ for $d/a_{\rm B}=2.2,2.7$ and $3.5$ respectively.     

\section*{Discussion}
We performed atomic resolution measurements resolving spin-spin interactions of interacting dopants, realizing quantum simulation of a two-site Hubbard system. Analyzing these local measurements of the spectral function\cite{Fischer:2007fa}, we find increasing Coulomb correlations and entanglement entropy as the system is ``stretched''\cite{Wiseman:2003jx,Ghirardi:2004dl,Amico:2008en} in the regime of non-perturbative interaction strengths $\mathcal{U}/t$.  Our experiment is the first to combine low effective temperatures $t/T \sim 30$ at $4.2$ K and single-site measurement resolution, considered essential\cite{Esslinger:2010vf,Greif:2013kb,Georgescu:2014bg} to simulate emergent Hubbard phenomena\cite{Anderson:1987ii,Gull:2013hh}.  Lower effective temperatures $t/T \sim 420$ are possible at $T=0.3$ K. For example, $4\times4$ Hubbard lattices with $\mathcal{U}/t=4\rightarrow 7$ and $t/T \sim 40$ have recently been associated with both the pairing state and pseudogap in systems exhibiting unconventional superconductivity\cite{Gull:2013hh}.  

The approach generalizes to donors, which can be placed in silicon with atomic-scale precision\cite{Fuechsle:2012bl} and spatially measured \textit{in-situ} after epitaxial encapsulation\cite{Miwa:2013ib,Saraiva:2016bs}.  In contrast to disordered systems\cite{Richardella:2010iv}, atomically engineered dopant lattices will require weak coupling to a reservoir, displaced either vertically as demonstrated herein, or a laterally\cite{Fuechsle:2012bl}.  Strain could be used to further enhance the splitting between light and heavy holes, or suppress valley interference processes of electrons\cite{Koiller:2002ih,Salfi:2014kaa}.  Interestingly, open Hubbard systems which may exhibit unusual Kondo behaviour\cite{Lopez:2002gz,Agundez:2015ey} could also be studied by this method. The demonstrated measurement of spectral functions could be used to directly determine excitation spectra, evaluate correlation functions\cite{Richardella:2010iv}, or obtain quasi-particle interference spectra\cite{Fischer:2007fa}, all of which contain rich information about many-body states, including charge-ordering effects. We envision \textit{in-situ} control of filling factor\cite{Anderson:1987ii,Gull:2013hh}, using a back-gate or patterned side-gate\cite{Fuechsle:2012bl}.  These capabilities will allow for quantum simulation of chains, ladders, or lattices\cite{Anderson:1987ii,Gull:2013hh,Dagotto:1996jl} at low effective temperatures, having interactions that are engineered atom-by-atom.

\section*{Acknowledgements}

We thank H. Wiseman, M. A. Eriksson, M. S. Fuhrer, O. Sushkov, D. Culcer, J.-S. Caux, B. Reulet, G. Sawatzky, J. Folk, F. Remacle, M. Klymenko and B. Voisin for helpful discussions.  This work was supported by the European Commission Future and Emerging Technologies Proactive Project MULTI (317707), the ARC Centre of Excellence for Quantum Computation and Communication Technology (CE110001027), and in part by the U.S. Army Research Office (W911NF-08-1-0527) and ARC Discovery Project (DP120101825). S.R. acknowledges a Future Fellowship (FT100100589). M.Y.S. acknowledges a Laureate Fellowship.  The authors declare no competing financial interests.

\section*{Author Contributions}
Experiments were conceived by J.S, J.A.M, and S.R.  J.S. carried out the experiments and analysis, with input from J.A.M., R.R., L.C.L.H, and S.R. Theory modeling was carried out by J.S., J.A.M., R.R., and L.C.L.H and S.R., with input from all authors. J.S. and S.R. wrote the manuscript with input from all authors. \\\\
$^*$ j.salfi@unsw.edu.au,$^\dagger$ s.rogge@unsw.edu.au

\section*{Methods}
\noindent{\bf Sample Preparation} Samples were prepared by flash annealing a boron doped ($p\approx10^{19}$ cm$^{-3}$) silicon wafer at $\sim 1200$ $^\circ$C in UHV followed by slow cooling at a rate $1$ $^\circ$C$\cdot$min$^{-1}$ to 340 $^\circ$C.  Then, hydrogen passivation was carried out $\sim 340$ $^\circ$C for ten minutes by thermally cracking H$_2$ gas at a pressure $P_{\rm H_2}=5\times10^{-7}$ mbar. \\

\noindent{\bf Measurements} Atomic resolution single-hole tunneling spectroscopy was performed at 4.2 K using an ultra-high vacuum Omicron low temperature scanning tunneling microscope (LT-STM).  Current $I$ was measured as a function of sample bias $U$ and ${\rm d}I/{\rm d}U$ was obtained by numerical differentiation. Details for the analysis of the data are provided in Figs.~\ref{fdIdUfit}, \ref{fSplitting}, \ref{fDepth}, \ref{fSingleBoron}, \ref{fDoubleBoron},\ref{fCorr},and \ref{fIVZ}, and Appendices \ref{sMeasuredSpectrum},\ref{sDepth},\ref{sSpatialModel} and \ref{sTipHeightSpectra}.\\

\noindent{\bf Theory} Theory calculations of interacting states were carried out using the configuration interaction approach, in the Luttinger-Kohn representation including a realistic description of the heavy-hole ($J=3/2$,$|m_J|=3/2$), light-hole ($J=3/2$,$|m_J|=1/2$), and split-off hole ($J=1/2$,$|m_J|=1/2$) degrees of freedom.  Details for the theory are provided in  Fig.~\ref{fTheoryQPWF} and Appendices \ref{sTheory},\ref{sHubbard} and \ref{sEntanglement}.

\section*{Supplementary Information}
\subsection{Tunneling spectroscopy}
\label{sMeasuredSpectrum}

\begin{figure}
\includegraphics{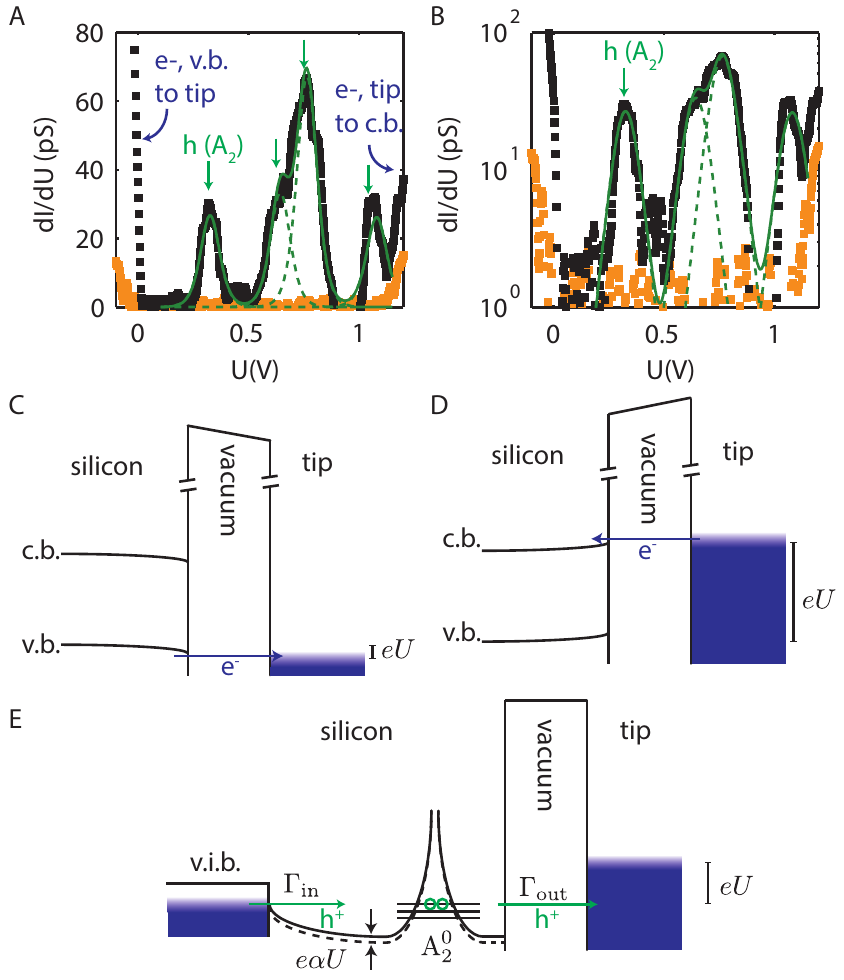}
\begin{flushleft}
\caption{A. ${\rm d}I/{\rm d}U$ vs. $U$ for the $d/a_{\rm B}=2.2$, above (black squares) and away from (orange squares) the coupled acceptors.  Least square fits (green line) considering sequential hole transport through several peaks (green dashed lines).  B.  Same as A on logarithmic scale.  C. Tunneling of electrons from valence band (v.b.) to the tip.  D. Tunneling of electrons from the tip to the conduction band (c.b.).  Schematic of hole tunneling at positive sample bias $U$, from valence impurity band reservoir, to acceptor pair, to tip.  Ground and excited states for two-hole occupation are shown.  Electrons (holes) in the hole reservoir and STM tip are shaded blue (white).  }
\label{fdIdUfit}
\end{flushleft}
\end{figure}

In this section we discuss the extraction of the energies of the states observed in ${\rm d}I/{\rm d}U$ tunneling spectra.  We find that the the number of observed states,  the state's relative energies to eachother, and their energies relative to the sample's Fermi energy, are only compatible with two interacting holes on pairs of acceptors.  We further estimate the total tunnel coupling $\Gamma_{\Sigma}$ to the tip and sample reservoirs for sequential hole transport, based on the ${\rm d}I/{\rm d}U$ lineshape, and show that $h\Gamma_{\Sigma}$ is much less than the inter-acceptor tunnel coupling $t$ for separations $d$ considered in the main text.  

For the acceptor pair with $d/a_{\rm B}=2.2$ (Fig.~\ref{fig3}A, main text), measured ${\rm d}I/{\rm d}U$ is plotted in linear scale (Fig.~\ref{fdIdUfit}A) and log scale (Fig.~\ref{fdIdUfit}B).  Away from the acceptor pair (Fig.~\ref{fdIdUfit}A/B, orange squares), direct tunneling of electrons from the valence band to the tip (holes from the tip to the valence band) is obtained for $U\lesssim 0$ V as illustrated in Fig.~\ref{fdIdUfit}C, while direct tunneling of electrons from the tip to the conduction band occurs for $U\gtrsim 1.15$ V as illustrated in Fig.~\ref{fdIdUfit}D.  Above the pair (Fig.~\ref{fdIdUfit}A/B, black squares), we observe a well-isolated peak at $U\approx 0.3$ V, a collection of two strong (closely spaced) peaks at $0.6$ V and $0.75$ V, and fourth peak at $U\approx 1.0$ V.  

Plotted on a logarithmic scale, both the current (not shown) and numerically differentiated ${\rm d}I/{\rm d}U$ data (Fig.~\ref{fdIdUfit}B), black squares) grow exponentially with increasing $U$ for the first peak before reaching a local maximum.   As suggested by this observation\cite{Foxman:1993jq,Mol:2013dj}, we fit the first peak of the differential conductance to a state probed by sequential tunneling with broadening determined by the $T=4.2$ K temperature of the reservoir, $\textrm{cosh}^{-2}(e \alpha (U-U_1)/2k_BT)$, where $e=1.602\times10^{-19}$ C is the elementary charge, $k_B=1.381\times10^{-23}$ J/K is the Boltzmann constant, $U_1$ is the peak voltage, and $\alpha$ is the well-known lever arm from sequential transport\cite{Foxman:1993jq} describing the variation in the energy level $\Delta E = e\alpha \Delta U$ due to a change in applied bias $\Delta U$, as schematically illustrated in Fig.~\ref{fdIdUfit}E.  

The least-squares fit is in good agreement with the measured ${\rm d}I/{\rm d}U$ for both the full-width at half maximum $\Delta U=3.6k_BT/e\alpha$ and the exponential decay of the tail $\exp(e\alpha (U-U_1)/k_BT)$.  A peak voltage $U_1=334.0 \pm 0.9$ meV and a lever arm $\alpha=0.0144\pm0.0006$ were obtained for tunneling of holes from the reservoir, to the acceptor pair, to the tip.  The value for $\alpha$ is similar to those found for single-hole transport through individual acceptors in flashed p-type silicon\cite{Mol:2013dj} and single arsenic donors in flashed n-type silicon\cite{Salfi:2014kaa}.  This excellent agreement shows that broadening of the single-hole transport peaks is due to the temperature associated with the Fermi-Dirac distribution in the hole reservoir. 

The small value of $\alpha$ means that the ``broadened'' lineshape in Fig.~\ref{fdIdUfit}A and Fig.~\ref{fdIdUfit}B is due to weak electrostatic control of the sample potential by the STM tip, as expected due to strong screening by the carrier reservoir.  This is further confirmed by the negligible ``spectral shift''\cite{Lin:1998hh}, that is, the similarity of the apparent gap $\approx 1.15$ eV in ${\rm d}I/{\rm d}U$ measurements of Fig.~\ref{fdIdUfit}A/B to the actual $1.14$ eV band gap for heavily-doped p-type silicon at low temperature\cite{Wagner:1988dpa}.

\subsubsection{Energy spectrum}

\begin{figure}
\includegraphics{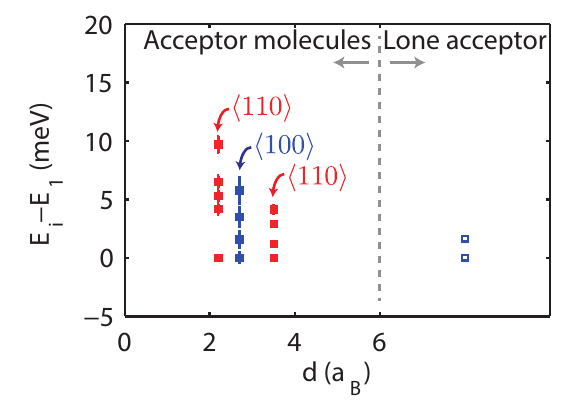}
\begin{flushleft}
\caption{Extracted energies $E_{i}-E_{1}$ of observed states relative to the lowest energy state (energy $E_1$), for $d/a_{\rm B}=2.2$, $d/a_{\rm B}=2.7$, and $d/a_{\rm B}=3.5$.  For comparison, the energies of the predominantly $\pm$``3/2'' and $\pm$``1/2'' spin Kramers doublets of a lone acceptor, $\sim 1$ nm from the Si[001]:H interface, are shown on the right hand side.  }
\label{fSplitting}
\end{flushleft}
\end{figure}

We extracted the lever arm $\alpha$ and energies of states $E_i-E_{1}=e\alpha(U_i-U_1)$, by least-squares fitting of ${\rm d}I/{\rm d}U$ to a sum of single-hole tunneling peaks\cite{Mol:2013dj,vanderHeijden:2014fp,Salfi:2014kaa,Voisin:2015gl,Mol:2015im}.  Fits to ${\rm d}I/{\rm d}U=\sum_{i}A_i(z)\big(\textrm{cosh}^{-2}(e \alpha (U-U_i)/2k_BT)\big)$ are found in  Fig.~\ref{fdIdUfit}A and \ref{fdIdUfit}B (solid green lines) along with individual peaks (dashed green lines), on linear and log scales, respectively.  Extracted energies $E_i-E_1=e \alpha (U_i-U_1)$ of the excited states are plotted in Fig.~\ref{fSplitting}, relative to the energy $E_1$ of the two-hole ground state, for $d/a_{\rm B}=2.2$, $d/a_{\rm B}=2.7$ and $d/a_{\rm B}=3.5$.  

For all $d$ the extracted energy splittings between the states (Fig.~\ref{fSplitting}) are too small to be associated with single-hole occupied states (A$_2^-$ states) of coupled acceptors.  For $d/a_{\rm B}=2.2$, two energetically nearby states were found at $5$ and $6.5$ meV, while a higher energy state was found at $\approx 3.5$ meV above those.  Here, the observed splittings are incompatible with the ``1s'' hybridization energy $2t \approx 25$ meV between tunnel-hybridized single-hole states estimated by numerics (Section \ref{sTheory}).  Similar arguments hold for the $d/a_{\rm B}=2.7$ ($d/a_{\rm B}=3.5$), where the excited state splittings $\approx 2$ meV ($\approx 1$ meV) are incompatible with the single-hole coupling energy scale of $2t\approx 14$ meV ($2t \approx 7$ meV).  

Even more importantly, it follows from the electrostatic arguments given below that the coupled-acceptor peaks observed in experiments are too close to the hole reservoir's chemical potential to correspond to ionizations of the very ``deep'' single-hole (A$_2^{-1}$) states into zero-hole (A$_2^{-2}$) states.  Rather, they are only compatible with ionizations of the two-hole (A$_2^0$) ground and excited states of acceptor pairs into one-hole (A$_2^{-1}$) states, which are energetically very similar to the ionization energy of a neutral (A$^0$) acceptor.  For $d/a_{\rm B}=2.2\rightarrow3.5$ we found ionization transitions A$_2^{-1}\rightarrow$A$_2^{-2}$ for the last hole (Section \ref{sTheory}) require more than $100$ meV energy.  This exceeds the Fermi energy $E_F-E_V\approx 50$ meV of the reservoir ($E_V$ is the valence band edge) by more than $50$ meV.  Now, for $U=0$ V, the tip's contact potential bends states by an amount $-e\alpha U_{\rm FB}=-e\alpha (W-\Phi_{\rm tip})$ relative to the sample reservoir, where $W=5.2$ eV is the work function of degenerately doped p-type reservoir, and $\Phi_{\rm tip}$ is tungsten tip workfunction.  Let us assume a work function $\Phi_{\rm tip}=4.8\pm 0.3$ eV typical of a tungsten tip\cite{Mol:2013dj}.  Then, the states bend down by $5.8\pm 4.2$ meV at zero bias, much smaller than the value $>50$ meV required to depopulate a single hole A$_2^-$ state (A$_2^{-1}\rightarrow$A$_2^{-2}$ transition).  Therefore at zero bias, the observed acceptor pairs are in an A$_2^-$ charge state, and a bias $U=\alpha^{-1}\times 50$ mV much less than $-1$ V would be required to remove the last hole by tip-induced band bending.  This is incompatible with $U\sim 0.2$ to $0.3$ V of peaks observed in experiments.  On the other hand, ionization energies of the two-hole states, A$_2^0\rightarrow$A$_2^-$ of coupled acceptors are calculated (Section \ref{sTheory}) to be very energetically similar to A$_0$ ionization energies.  Consequently, transitions populating ground and excited two-hole will occur for $U>0$, explaining how they (much like single-acceptor ionizing resonances\cite{Mol:2013dj}) are observed in our experiments.  

\subsubsection{Transport coupling to reservoir}
\label{sCouplingEstimate}

\setlength{\tabcolsep}{5pt}
\begin{table}
\begin{tabular} {| l | l | l | l | l |}
\hline
$d/a_{\rm B}$ & Fit \# 1 & & Fit \#2 & \\
\hline
& $|\gamma_{\rm oo}|$ & $\overline{SSe}$ & $|\gamma_{\rm oo}|$ & $\overline{SSe}$ \\
\hline
2.2 & free & 0.0159 & fixed,0 & 0.0293 \\
2.7 & free & 0.0126 & fixed,0 & 0.0509 \\
3.5 & free & 0.0391 & fixed,0 & 0.0638 \\
\hline
\end{tabular}
\caption{Comparison of sum of square errors when correlation parameter $\gamma_{\rm oo}$ is a free parameter in fitting model (Fit \# 1), versus when $\gamma_{\rm oo}$ is fixed to zero (Fit \#2).}
\label{tSSE}
\end{table} 

\setlength{\tabcolsep}{5pt}
\begin{table}
\begin{tabular} {| l | l | l | l |}
\hline
$d(a_{\rm B})$ & $2.2$ & $2.7$ & $3.5$\\
\hline
\hline
$t$ (GHz) & 2900 & 1700 & 850 \\
$h\Gamma_{\Sigma}$ (GHz) & $<2.5$ & $<30$ & $<10$ \\
\hline
$t/h\Gamma_{\Sigma}$ & $>1200$ & $>60$ & $>85$ \\
\hline
\end{tabular}
\caption{Estimated tunnel coupling $2t$ and environmental coupling rates $\Gamma_{\Sigma}$  for each $d$.  Conservative estimates are given for the molecular coupling, based on calculations for the [100] direction.  Nevertheless, even the conservative molecular coupling estimate for the $[100]$ direction well exceed {\it upper bounds} on the environmental coupling, estimated from the ${\rm d}I/{\rm d}U$ lineshape. }
\label{tCouplingTable}
\end{table}

We have used the ${\rm d}I/{\rm d}U$ lineshape to obtain an upper bound on the tunnel coupling $\Gamma_{\Sigma}=\Gamma_{\rm in}+\Gamma_{\rm out}$ to the environmental reservoirs - the largest lifetime broadening energy scale\cite{Foxman:1993jq} $h\Gamma_{\Sigma}$ that can be included in the lineshape analysis of the data\cite{Mol:2013dj}.  As necessary to observe molecular states, we find $\Gamma_{\Sigma}$ for data in Fig.~\ref{fig3}A, \ref{fig3}B, and \ref{fig3}C, to be considerably smaller than the inter-acceptor tunnel coupling $t$ estimated from the hybridization energies (Table \ref{tCouplingTable}).  Tip height-dependent measurements (Section \ref{sTipHeightSpectra}) confirmed the exponential dependence of the tunnel current on tip height, indicating the rate to the tip is the slower (dominant) rate.  The dwell-time (and coupling $\Gamma_{\Sigma}$), determined by the fast rate, is therefore controlled by the coupling to the hole reservoir in the substrate, $\Gamma_{\rm in}$ in Fig.~\ref{fdIdUfit}E.  

\subsection{Acceptor depth estimate}
\label{sDepth}

We have estimated the depth below the silicon surface of the acceptor dopant pairs, presented in Fig.~\ref{fig3}A, \ref{fig3}B and \ref{fig3}C of the main text, to be approximately $0.5$ nm, $0.9$ nm, and $0.9$ nm, respectively.  The depth of isolated dopants in scanning tunneling spectroscopy is typically estimated fitting the spatial variation of a spectral feature such as a band edge, in a bias condition where the dopant is ionized\cite{Teichmann:2008bh, Lee:2010ko, Mol:2013dj, Salfi:2014kaa}, to a dielectric screened Coulomb potential.  However, as illustrated in Fig.~\ref{fdIdUfit}E, tunneling from the reservoir to the acceptor pair is a transition from a one-hole state to a two-hole state.  Consequently, for voltages $U$ below the lowest voltage peak at $U=U_1$ in Fig.~\ref{fdIdUfit}A and Fig.~\ref{fdIdUfit}B, the two-acceptor system is in Coulomb blockade with single-hole occupation, rather than being doubly ionized, such that fitting to dielectric-screened single-ion potentials is inappropriate\cite{Teichmann:2008bh, Lee:2010ko, Mol:2013dj, Salfi:2014kaa}.  For $U$ above $U_1$ the two-acceptor system has a high probability of occupation by two holes since $\Gamma_{\rm out} \ll \Gamma_{\rm in}$.  Hence, the method of fitting the band profile to dielectric screened ion potentials used in the above references is problematic.

To estimate the depth of our acceptors, we use the well-known result that the spatial extent of the bound state measured by STS increases with increasing dopant depth (see \textit{e.g.}, reference \onlinecite{Celebi:2010hg}).  We compare the full-width at half-maximum $W$ of the spatial tunneling distribution fit in Section \ref{sSpatialModel} to the same quantity calculated for subsurface acceptor-bound holes in the $6\times6$ spin-orbit coupled valence band.  We provide this estimate for completeness, since the fitting procedure described in Section \ref{sSpatialModel} does not rely on prior knowledge of the depth or Bohr radius of the dopant.

\begin{figure}
\includegraphics{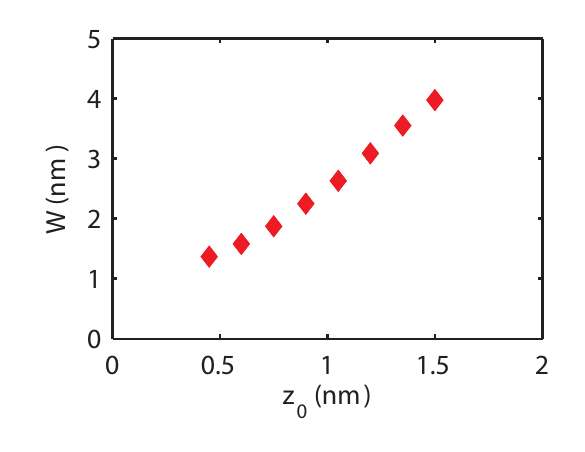}
\begin{flushleft}
\caption{Dependence of the full-width half-maximum $W$ of the spatial tunneling probability distribution on depth of acceptor beneath the surface, calculated by numerical diagonalization of the $6\times6$ spin-orbit coupled Kohn-Luttinger Hamiltonian.}
\label{fDepth}
\end{flushleft}
\end{figure}

The full-width at half maximum $W$ for the 1s orbital density $\phi_s(\mathbf{r})\sim \exp(-|\mathbf{r}-\mathbf{r}_0|/a*)$ at $z=0$ can be directly solved giving $W=2\sqrt{(\tfrac{1}{2}a^*\log(\tfrac{1}{2})-z_0^*)^2-(z_0^*)^2}$.  Evaluating this quantity we obtain $W=1.48\pm 0.12$ nm, $W=2.25\pm0.39$ nm, and $W=2.19\pm0.25$ nm for the measured molecules with $d/a_{\rm B}=2.2$, $2.7$, and $3.5$, respectively.  Comparing these results to the full-width at half maximum $W$ predicted for the orbitals using the numerical solution to Kohn-Luttinger $6\times6$ Hamiltonian presented in Fig.~\ref{fDepth}, the $W$ extracted from experiments translate to  depth estimates of $\approx 0.53$ nm, $\approx 0.9$ nm, and $\approx 0.9$ nm, respectively.  

\subsubsection{Correlation with lever arms}
When the acceptors are located deeper in the sample, the lever arms are anticipated to be smaller\cite{Mol:2013dj}.  The relative depths of the planes in which the acceptor pairs are estimated to be located agree with the extracted lever arms.  We obtain a lever arm $\alpha=0.0144\pm0.0006$ for the shallower $d/a_{\rm B}=2.2$ pair.  The other two $d/a_{\rm B}=2.7$ and $d/a_{\rm B}=3.5$ pairs which are slightly deeper have lever arms $\alpha = 0.0088\pm 0.0010$ and $\alpha=0.0055\pm0.0003$, respectively.  We note that fluctuations in the depth of the reservoir can also influence the lever arm.  We anticipate that varying the dopant depths has a small effect on correlations in our experiments where the overall system is neutral.  We have observed that the neutral level of near-surface acceptors ($N=1$) remains bulk-like\cite{Mol:2013dj} for depths 0.5-2.0 nm, as expected, due to a competition between the confinement and the dielectric mismatch.  Fixed ionization energy essentially fixes the Bohr radius, fixing $t$ and $U/t$. We note however that the apparent spatial extent of the wavefunction increases with increasing depth.  This is not because of a change in the physical Bohr radius.  Rather, it occurs because STM probes the wavefunction near the surface. 

\subsection{Theory of interacting acceptors}
\label{sTheory}

We employed a microscopic configuration interaction framework to calculate eigenstates of interacting holes bound to proximate acceptors, in a Kohn-Luttinger $6\times6$ spin-orbit coupled representation (heavy holes, light holes, and split-off holes), and in a simplified single-band representation for reference.  In both representations, the two-hole ground state at relatively small $d/a_{\rm B}\sim2$ was found to be predominantly composed of the $\left|\rm{e}+ \rm{e}-\right\rangle$ singlet of two even orbitals, where $+$ and $-$ denote +``3/2'' and $-$``3/2'' respectively for the valence band holes, and denote $\uparrow$ and $\downarrow$ respectively in the single-band approximation.  The probability amplitude of the $\left|\rm{o}+ \rm{o}-\right\rangle$ singlet of odd orbitals was found to increase with increasing $d$, signaling interaction-driven generation of quantum correlations.  The corresponding spatial tunneling probability distribution was predicted using the quasi-particle wavefunction (QPWF) density appropriate for few-particle states with quantum correlations\cite{Rontani:2005eb,Maruccio:2007kl}, and are presented herein for the non-trivial $6\times6$ spin-orbit coupled case. 

\subsubsection{Overview}

In the absence of spin-orbit coupling, wavefunctions of interacting spins can be classified into singlets and triplets\cite{DC:1981up}.  For holes in the valence band, the atomic orbitals have p-like symmetry, and the angular momentum $L=1$ couples to the spin (intrinsic) angular momentum $S=1/2$, such that the resulting Bloch states have definite $J$ and $J_z$.  While intrinsic angular momentum is not a good quantum number, spin can nevertheless be generalized to hole pseudospin, \textit{i.e.}, Kramers doublets of time reversal symmetric linear combinations of Bloch states\cite{Kavokin:2004cd}.  

Coupling of hole spins is well understood in the Kohn-Luttinger framework\cite{Kavokin:2004cd,Climente:2008cx,Yakimov:2010fy}.  It has been theoretically shown that coupled hole pseudospin interact to produce singlets and triplets\cite{Kavokin:2004cd}, as for coupled electron spins.  In the Kohn-Luttinger approach generalized valence band holes spins as written as spinors $\phi_i(\mathbf{r}) = \sum_{J,m_J}F^i_{J,m_J}(\mathbf{r})|J,m_J\rangle$, where $F^i_{J,m_J}(\mathbf{r})$ are the envelope functions corresponding to the Bloch states $\left|J,m_J\right\rangle$.  Due to spin-orbit coupling such spinors are $\sim 90$ \% polarized (rather than 100 \% polarized) into single Bloch components.  This produces a small band-mixing effect when the pseudospins are coupled together\cite{Climente:2008cx,Yakimov:2010fy}, that we take into account exactly in our numerics.  

\subsubsection{Full-Configuration Interaction}
\label{sTheoryMethodology}

We employ the Kohn-Luttinger FCI approach developed for valence band holes in reference [\onlinecite{Yakimov:2010fy}].  Two-hole eigenstates $\left|\Psi\right\rangle$ are expressed as superpositions of antisymmetrized two-hole Slater determinants $\left|\alpha\beta\right\rangle=c^\dagger_{\alpha}c^\dagger_{\beta}|0\rangle=2^{-1/2}(\left|\alpha\right\rangle_1\left|\beta\right\rangle_2-\left|\beta\right\rangle_1\left|\alpha\right\rangle_2)$ with probability amplitudes $d_{\alpha\beta}$,
\begin{equation}
|\Psi\rangle=\sum_{\alpha,\beta}d_{\alpha\beta}\left|\alpha\beta\right\rangle.
\end{equation}
In $|\alpha\beta\rangle$ above, $c_j^\dagger$ creates a single-hole state $|j\rangle=c^\dagger_j|0\rangle$, where $|0\rangle$ is the vacuum state.  As discussed in detail below, the single-particle states $\left|j\right\rangle$ were chosen as the eigenstates of the non-interacting Hamiltonian $\mathcal{H}_0$ of a pair of acceptor ion potentials and a hard-wall interface with dielectric mismatch.  The Hamiltonian for the two-hole system was taken as $\mathcal{H}=\mathcal{H}_0(\textbf{r}_1)+\mathcal{H}_0(\textbf{r}_2)+V_{\rm h-h}(\textbf{r}_1,\textbf{r}_2)$ where $V_{\rm h-h}(\textbf{r}_1,\textbf{r}_2)$ is the Coulomb repulsion term for the holes. Substituting $\left|\Psi\right\rangle$ above into $\mathcal{H}|\Psi\rangle=E|\Psi\rangle$ and taking the inner product with $\left\langle \nu\lambda\right|$, we obtain
\begin{widetext}
\begin{equation}
(E_{0\nu}+E_{0\lambda})d_{\nu\lambda} + J_{\nu\lambda} - K_{\nu\lambda} + \sum_{\alpha\beta}d_{\alpha\beta}(1-\delta_{\alpha\beta,\nu\lambda})(\Gamma_{\nu\lambda,\alpha\beta}-\Gamma_{\nu\lambda,\beta\alpha}) = E d_{\nu\lambda},
\label{eq:ci}
\end{equation}
\begin{equation}
 \Gamma_{\nu\lambda,\alpha\beta}=\int d\textbf{r}_1 d\textbf{r}_2 \Big(\sum_{J,m_j} F^*_{\nu,J,m_j}(\textbf{r}_1)F_{\alpha,J,m_j}(\textbf{r}_1)\Big) V_{\rm h-h}(\textbf{r}_1,\textbf{r}_2)\Big(\sum_{J',m_j'}F^*_{\lambda,J',m_j'}(\textbf{r}_2)F_{\beta,J',m_j'}(\textbf{r}_2)\Big),
\end{equation}
\end{widetext}
for each $\left|\nu\lambda\right\rangle$, where $\Gamma_{\nu\lambda,\alpha\beta}=\langle\nu|_1\langle\lambda|_2V_{\rm h-h}|\alpha\rangle_1|\beta\rangle_2$ is an off-diagonal (interaction) term evaluated in the spinor representation, $J_{\nu,\lambda}=\Gamma_{\nu\lambda,\nu\lambda}$ is the direct Coulomb interaction, and $K_{\nu\lambda}=\Gamma_{\nu\lambda,\lambda\nu}$ is a direct exchange interaction.  Finite $\Gamma_{\nu\lambda,\alpha\beta}$ terms can introduce correlations\cite{Yakimov:2010fy} by creating eigenstates that are admixtures of two-particle Slater determinants $\left|\alpha\beta\right\rangle$.

The single particle Hamiltonian employed was $\mathcal{H}_0=\mathcal{H}(\textbf{k})+V_{\rm{ion},A}+V_{\rm{ion},B}+V_{\rm{image},A}+V_{\rm{image},B}+V_{\rm{wall}}$ taking into account ion potentials $V_{\rm{ion},A/B}$ of both acceptors, image charges $V_{\rm{image},A/B}$ due to dielectric mismatch\cite{Hao:2009eq}, and an infinite (hard-wall) $V_{\rm{wall}}$ potential a distance $1a_{\rm B}\sim1$ nm away from the ion representing the semiconductor/vacuum interface.  For $\mathcal{H}(\mathbf{k})$ we considered both the $6\times6$ spin-orbit coupled Kohn-Luttinger valence band\cite{Luttinger:1955ee,Chao:1992fv} $\mathcal{H}_{\rm KL}(\mathbf{k})$, as well as a single-band, isotropic, parabolic dispersion relation $\mathcal{H}_{p}(\mathbf{k})=\hbar^2k^2/2m^*$ where $m^*$ is an effective mass chosen to reproduce boron's binding energy\cite{Belyakov:2007jf}.  The latter allows us to calculate states of a simple ``scaled hydrogen molecule'' without complex spin-orbit coupling and band structure effects. The ion potentials were taken as dielectric-screened Coulomb potentials.  For the $6\times6$ scheme we included additional charge $\delta q=-0.1e$ at the ion site as a crude core-correction to obtain the correct spatial extent $\left\langle r\right\rangle$ and ionization energy of the states\cite{Bernholc:1977bf}.  The interaction term $V_{\rm h-h}(\mathbf{r}_1s,\mathbf{r}_2)$ is a  dielectric screened Coulomb interaction including an image charge distribution produced by dielectric mismatch\cite{Orlandi:2001gg}.  

For the single band theory, the four lowest single-electron eigenstates of $\mathcal{H}_0$ were used to construct 6 possible two-particle configurations, all of which were included for a full configuration interaction (FCI) approach.  We considered two even, spin-degenerate orbitals $\left|{\rm e}\sigma\right\rangle=c^\dagger_{{\rm e}\sigma}\left|0\right\rangle$, and two odd, spin-degenerate orbitals $\left|{\rm o}\sigma\right\rangle=c^\dagger_{{\rm o}\sigma}\left|0\right\rangle$.  

For the $6\times6$ Kohn-Luttinger calculations, we use a basis that reflects the four low-energy states of single acceptors with s-like envelopes - the predominantly $|m_J|=3/2$ (``3/2'') state, and the predominantly $|m_J|=1/2$ (``1/2'') state\cite{Bir:1963iy}.  Near an interface, the ``1/2'' state experiences more confinement than the ``3/2'' state, making the ``3/2'' state the ground state\cite{Gammon:1986ks,Masselink:1985cs,Mol:2015im}.  For lone subsurface acceptors $\sim 1$ nm from an Si[001]:H interface, we have typically measured $\sim1-2$ meV splittings (see \textit{e.g.}, lone acceptor data in Fig.~\ref{fSplitting}), in good agreement with few meV values predicted by $6\times6$ Kohn-Luttinger.  See reference [\onlinecite{Mol:2015im}] for further details.  In the two-acceptor potential $\mathcal{H}_0$ these hybridize into eight single-hole eigenstates, which were used to construct 28 possible two-hole configurations, all of which were included in the state vector for a FCI approach.  For the acceptor distances in the main text, four were predominantly composed of heavy holes (``3/2'' states) and four were predominantly composed of light holes (``1/2'' states).  In both groups of four states, two are found within a Kramers doublet with even parity symmetry for the majority spin, and two are found within a Kramers doublet having odd parity symmetry for the majority spin\cite{Climente:2008cx,Yakimov:2010fy}.  

A numerical representation of the FCI matrix (Equation \ref{eq:ci}) was directly diagonalized to obtain state vectors $\textbf{d}$ and energies $E$.  The FCI matrix was obtained by evaluating the Coulomb matrix elements $J_{\nu\lambda}$, exchange matrix elements $K_{\nu\lambda}$, and the interaction terms $\Gamma_{\nu\lambda,\alpha\beta}$ by Monte-Carlo integration, using the VEGAS method for adaptive sampling and $10^7$ iterations per integral.  The basis of single particle eigenstates required for the integrals was obtained numerically using the finite difference scheme, for both the single-band and $6\times6$ representations, on a real-space grid of $110\times110\times110$ points with a discretization step size of $0.21$ nm, using Luttinger parameters for silicon's valence band\cite{Bernholc:1977bf}.  To normalize the measured and theoretical ($6\times6$ and single-band) inter-acceptor distances, we choose the value $a_{\rm B}=1.3$ nm reproducing the $44.4$ meV binding energy of a Boron acceptor in silicon\cite{Belyakov:2007jf} in a single-band approximation.  

\subsubsection{Correlations in ground state}
\label{sTheoryGroundState}

In this section, the ground state of the FCI Hamiltonian in Section \ref{sTheoryMethodology} is discussed.  Our FCI predicts a singlet ground state for both the spin-orbit coupled $6\times6$ representation and for the isotropic single-band case.  In a single-band representation (ignoring spin-orbit coupling), the ground state
\begin{equation}
|\Psi_S\rangle=(\gamma_{\rm ee} c_{\rm e,\uparrow}^\dagger c_{\rm e,\downarrow}^\dagger-\gamma_{\rm oo} c_{\rm o,\uparrow}^\dagger c_{\rm o,\downarrow}^\dagger)|0\rangle
\end{equation}
was obtained.  With increasing $d$, dynamical interactions ($\Gamma$'s in Equation \ref{eq:ci}) resulted in increasing $\gamma_{\rm oo}$ and increasing quantum correlations $C=2|\gamma_{\rm oo}|^2$, as predicted elsewhere\cite{Coulson:ty,He:2005in}.  

The ground state of interacting, spin-orbit coupled holes\cite{Climente:2008cx,Yakimov:2010fy} in the $6\times6$ Kohn-Luttinger representation contains contributions for which majority pseudospin have even parity,
\begin{equation}
\sum_{m_J \neq m'_J}d_{({\rm e},m_J),({\rm e},m'_J)}c_{{\rm e},m_J}^\dagger c_{{\rm e},m'_J}^\dagger\left|0\right\rangle,
\end{equation}
and contributions for which for majority pseudospin have odd parity,
\begin{equation}
\sum_{m_J \neq m'_J}d_{({\rm o},m_J),({\rm o},m'_J)}c_{{\rm o},m_J}^\dagger c_{{\rm o},m'_J}^\dagger\left|0\right\rangle.
\end{equation}
For small $d/a_{\rm B}\sim2$, the dominant configuration (approx.~$90$~\%) in the ground state singlet was found to be $d_{({\rm e},3/2)({\rm e},-3/2)}$, while for increasing $d$, the probability amplitude $d_{({\rm o},3/2)({\rm o},-3/2)}$ was found to increase.  From this, two conclusions can be drawn: (1) the increase in the probability amplitude $d_{({\rm o},3/2)({\rm o},-3/2)}$ with increasing $d$ signals the emergence of hole-hole quantum correlations\cite{Yakimov:2010fy}, and (2) the predominance of ``3/2'' pseudospinspin orbitals signals that the band-mixing effect is weak for the values of $d$ in experiments.  Contributions $d_{({\rm e},m_J),({\rm e},m_J')}$ and $d_{({\rm o},m_J),({\rm o},m_J')}$ with $m_J=\pm$3/2 and $m'_J=\pm$3/2 are much larger than the other ``band-mixing'' terms with $m_J=\pm$``1/2' or $m'_J=\pm$``1/2''.  

\subsubsection{Quasi-particle wavefunction density}

In this section, we discuss theoretical predictions of the correlation coefficient $C$ in Fig.~\ref{fig4}A, as well as theoretical calculation of the quantity measured for spatially resolved single-hole tunneling into multi-particle states with correlations, the QPWF density\cite{Rontani:2005eb,Maruccio:2007kl,Schulz:2015hv}.  The results of Section \ref{sTheoryGroundState} are considered, for both the single-band and spin-orbit coupled $6\times6$ Kohn-Luttinger representations.  

As described in Section \ref{sMeasuredSpectrum}, the tunneling probability density measured in experiments is determined by tunneling of a hole from the acceptor molecule to the tip, leaving behind a single-hole eigenstate $|i\rangle$ on the acceptor molecule.  The corresponding QPWF density $\Gamma^i_{\rm out}(\mathbf{r})\propto|M_{iS}(\mathbf{r})|^2$ is obtained\cite{Rontani:2005eb}, where $M_{iS}(\mathbf{r})=\langle i|\Psi(\mathbf{r})|\Psi_S \rangle$, $\Psi(\mathbf{r})=\sum_j c_j\phi_j(\mathbf{r})$ is the field operator, $\mathbf{r}=(x,y,z_0)$ is position of the tip apex orbital, and $\left|\Psi_S\right\rangle$ is the ground state.  For the single-hole transport processes probed, it is necessary to sum over the final (single-hole) states $\left|i\right\rangle$ of the molecule after hole tunneling to the tip.  The total tunneling probability density is then $\Gamma_{\rm out}(\mathbf{r})=\sum_i\Gamma^i_{\rm out}(\mathbf{r})$.  

For the single band case, evaluating the QPWF using the field operator gives $\Gamma_{\rm out}(\mathbf{r})\propto\left[|\gamma_{ee}|^2|\phi_e(\mathbf{r})|^2 + |\gamma_{oo}|^2|\phi_o(\mathbf{r})|^2\right]$.  Results for $C=2|\gamma_{oo}|^2$ are plotted in the main text (Fig.~\ref{fig4}A, dashed line).  

For the $6\times6$ spin-orbit coupled representation, the quasi-particle wavefunction densities are also readily obtained, such that $\Gamma_{\rm out}(\mathbf{r})=\sum_i|M_{iS}(\mathbf{r})|^2$, where $M_{iS}(\mathbf{r})=\sum_{\alpha\beta}d_{\alpha\beta}[\phi_\alpha(\mathbf{r})\delta_{i\beta}-\phi_b(\mathbf{r})\delta_{i\alpha}]$.  Evaluating this expression results in a QPWF density that is a sum of even and odd QPWF contributions for all $d$ and both orientations $\left\langle 100 \right\rangle$ and $\left\langle 110 \right\rangle$, as expected considering the generalized parity of holes\cite{Climente:2008cx}.  Results for $C$, again defined as twice the probability density of odd orbitals, are plotted in the main text (Fig.~\ref{fig4}A, coloured lines) for both $\left\langle 100 \right\rangle$ and $\left\langle 110 \right\rangle$ orientations.  

\begin{figure*}
\begin{flushleft}
\includegraphics{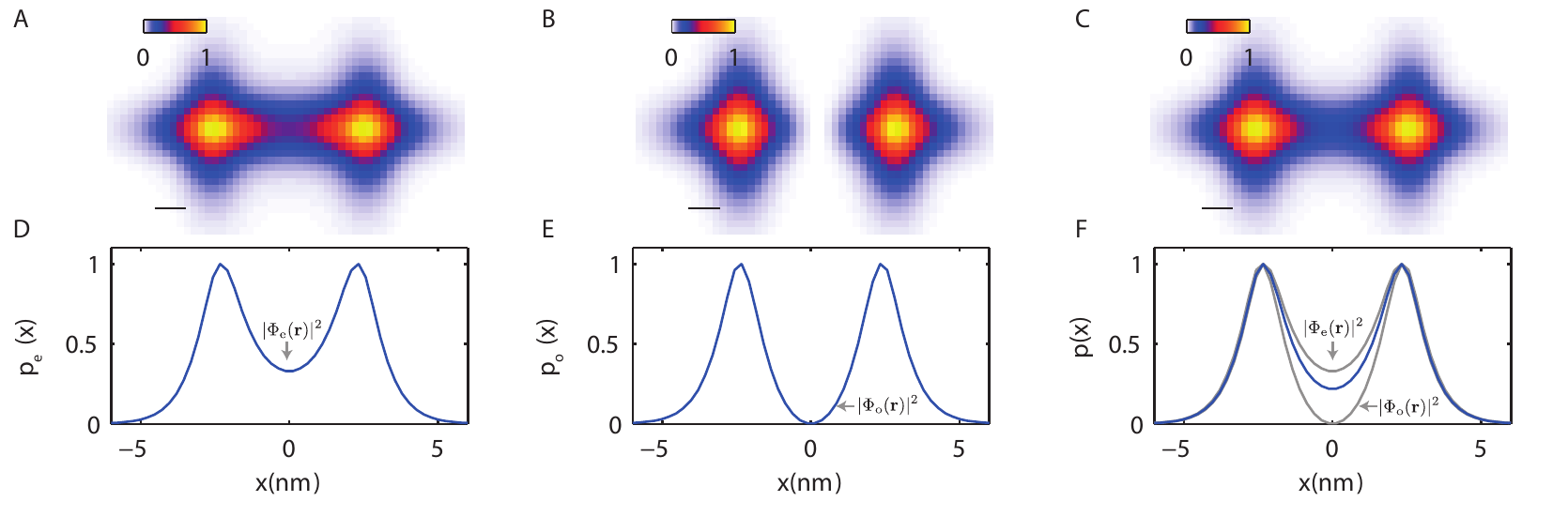}
\caption{A. Predicted spatial tunneling probability distribution for even quasiparticle wavefunctions using $6\times 6$ Kohn-Luttinger.  B.  Same as A, but for odd quasiparticle wavefunctions.  C.  Total tunneling probability distribution, a sum of A and B.  D.  Profile of spatial tunneling probability distribution for even quasiparticle wavefunctions using $6\times 6$ Kohn-Luttinger.  E.  Same as D, but for odd quasiparticles wavefunctions.  F.  Total tunneling probability distribution, a sum of of D and E.  Scale bar: 1 nm} 
\label{fTheoryQPWF}
\end{flushleft}
\end{figure*} 

The predicted spatial tunneling probability density is shown $1$ nm above the ion in Fig.~\ref{fTheoryQPWF} for an acceptor molecule with $d=3.8a_{\rm B}$.  The predicted correlation coefficient is $C=0.6$, similar to $C=0.78\pm0.16$ extracted from measurements of the $d=3.5a_{\rm B}$ acceptor molecule, in Fig.~\ref{fig4}A of the main text.  Contributions  to the spatial maps (line profiles) from even and odd QPWF densities are separately plotted in Fig.~\ref{fTheoryQPWF}A (Fig.~\ref{fTheoryQPWF}D) and Fig.~\ref{fTheoryQPWF}B (Fig.~\ref{fTheoryQPWF}E), respectively.  Their sum, the total tunneling probability density (line profile), is plotted in Fig.~\ref{fTheoryQPWF}C (Fig.~\ref{fTheoryQPWF}F).  Upper and lower grey lines in Fig.~\ref{fTheoryQPWF}F denote normalized QPWF densities for the even QPWF and odd QPWF components, respectively, which in this case directly correspond to the profiles in Fig.~\ref{fTheoryQPWF}D and \ref{fTheoryQPWF}E, respectively.

\subsection{Fitting of measured spatial tunnelling probabilities}
\label{sSpatialModel}

The experimentally obtained spatial tunneling probabilities for the two-hole ground state of acceptors, in Fig.~\ref{fig3}A, \ref{fig3}B, and \ref{fig3}C of the main text, were fit assuming even and odd linear combinations of atomic orbitals with s-like envelopes for $\phi_{e}(\mathbf{r})$ and $\phi_{o}(\mathbf{r})$. This data was obtained using a tip height established by constant current imaging conditions, at a bias where the topography was nominally flat apart from dimer row corrugations.

In this section we describe the fits and show that the probability density of an s-like wavefunction describes the measured probability density of the ground state and first excited state of an isolated acceptor, in agreement with what is expected for the s-like ``3/2'' pseudospin ground state and s-like ``1/2'' pseudospin excited state.

\begin{figure}
\includegraphics{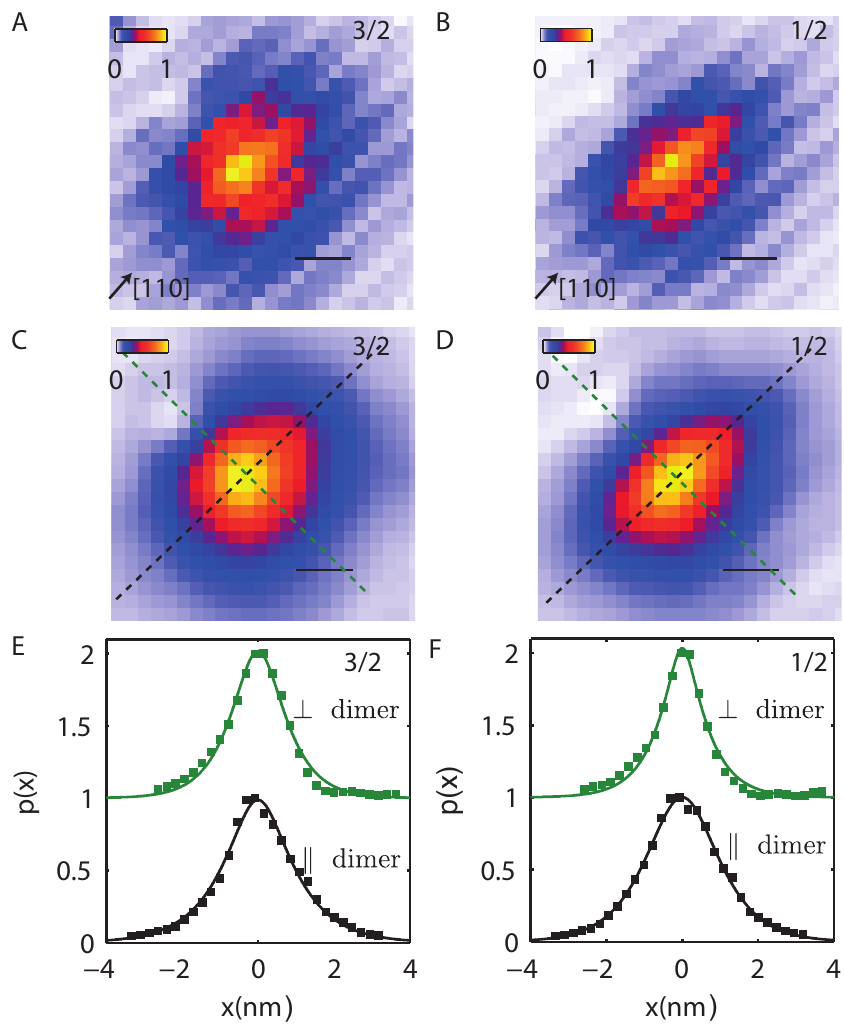}
\begin{flushleft}
\caption{A. Integrated ${\rm d}I/{\rm d}U$ for single acceptor ground state.  B.  Integrated ${\rm d}I/{\rm d}U$ for first excited state, $1.6\pm0.4$ meV above the ground state.  C.  Same as A with lattice frequencies filtered out.  D. Same as B with lattice frequencies filtered out.  E,F.  Line profile of relatively tunneling probability parallel (black squares) and perpendicular (green squares, offset for clarity) to the dimer, and fit of profile parallel (black line) and perpendicular (green line, offset for clarity) to dimer direction.  F. Same as E, but for first excited state. Scale bar: 1 nm}
\label{fSingleBoron}
\end{flushleft}
\end{figure}

Spatially resolved tunneling spectra were measured for isolated acceptors, consistently revealing two ${\rm d}I/{\rm d}U$ peaks associated with the s-like $J=3/2$ ground state manifold.  The $\sim2$ meV splitting of the isolated acceptor in Fig.~\ref{fSplitting} was found to be in good agreement with single-particle eigenstates calculated from our numerical Kohn Luttinger solver for depths determined as in previous work\cite{Mol:2013dj}, that is, by a fitting the profile to the ionized acceptor's perturbation to the band edge\cite{Teichmann:2008bh,Lee:2010ko}.  The spatial tunneling probability distributions for the lowest energy state and first excited state of an isolated acceptor were obtained by integrating the two low energy ${\rm d}I/{\rm d}U$ peaks.  Results are shown in Fig.~\ref{fSingleBoron}A and Fig.~\ref{fSingleBoron}B for the single acceptor in Fig.~\ref{fSplitting}.  

The dimer lattice structure of the hydrogen-terminated $2\times1$ surface is evident in the states Fig.~\ref{fSingleBoron}A and Fig.~\ref{fSingleBoron}B, running along the $[110]$ direction indicated.  The dimer and lattice frequencies were Fourier filtered out of the image\cite{Celebi:2008jm}, after which only the characteristic envelope of the probability density of the states remain.  Results for the ground state and first excited state are presented in Fig.~\ref{fSingleBoron}C and \ref{fSingleBoron}D, respectively.  The tunneling probability density reflects the hole density, and the appearance of two s-like envelopes is in agreement with expectations of the s-like envelopes for the ``3/2'' and ``1/2'' states\cite{Bir:1963iy,Mol:2015im}.  We observe that the measured low-frequency envelope for the probability density is closer to isotropic for the ground state, and spatially elongated along the dimer direction for the excited state.

The probability density was fit along a line in the plane ($z=0$) of the surface of the form $|\phi_s(\mathbf{r})|^2$, where $\phi_s(\mathbf{r})\sim \exp(-|\mathbf{r}-\mathbf{r}_0|/a*)$ is a 1s orbital envelope function.  In this expression, $\mathbf{r}=(x,y,z)$ is the tip position, $\mathbf{r}_0=(x_0,y_0,z_0*)$, $x_0$ and $y_0$ are centre coordinates of the orbital, $z^*_0$ is an effective depth, and $a^*$ is an effective Bohr radius.  The profile of the measured ground state (solid squares) and least-square fit (solid line) are shown for a profile along the direction parallel (black) and perpendicular (green) to the dimer row in Fig.~\ref{fSingleBoron}E, demonstrating excellent agreement.  The same quantities are plotted for the first excited state in Fig.~\ref{fSingleBoron}F.  A more noticeable anisotropy of the excited ``1/2'' state was found.  

Motivated by the theoretical observation that the QPWF density partitions into contributions from even and odd states, we categorize the measured QPWF density into contributions from even and odd QPWFs, that is, linear combinations of atomic orbitals with even and odd parity, respectively. For a distance $R$ between acceptors, we use $\phi_{e}(\mathbf{r})=(1/I_{e}(R))[\phi_s(\mathbf{r}-\mathbf{R}/2)+\phi_s(\mathbf{r}+\mathbf{R}/2)]$ and $\phi_{o}(\mathbf{r})=(1/I_{o}(R))[\phi_s(\mathbf{r}-\mathbf{R}/2)-\phi_s(\mathbf{r}+\mathbf{R}/2)]$, where $I_e(r)=1+(1+r+r^2/3)\exp(-r)$ and $I_o(r)=1-(1+r+r^2/3)\exp(-r)$ in atomic units.  In the fits presented in the main text, $a^*$, $z_0^*$, $\mathbf{R}$, and $|\gamma_{ee}|^2$ are free parameters, while $1=|\gamma_{oo}|^2+|\gamma_{ee}|^2$ is simply a consequence of normalization.  Effective Bohr radii $a^* \sim 1$ nm and depths $z_0^*\sim 1$ nm were obtained from the least squares fits.  In our fits the values for $a^*$ and $z_0^*$ define a full-width at half maximum $W$ for the spatial tunneling probability of each acceptor.  In Section \ref{sDepth}, the actual depth $z_0$ of the acceptors was estimated using $W$ derived from experimentally fit values $a^*$, $z_0^*$.

\subsubsection{Error analysis in fit}
The sum of square errors 
\begin{equation}
\overline{SSe}=\sum_{x_i}(\Gamma(x_i)-\Gamma_{\rm fit}(x_i))^2/(\Gamma(x_i))^2
\end{equation}
was used to determine the importance of the Coulomb correlation parameter $\gamma_{\rm oo}$ in the fits.  The first column (Fig \#1) of Table~\ref{tSSE} gives the results for the $\overline{SSe}$ when $|\gamma_{\rm oo}$ is a free parameter, while the second column (Fig \# 2) of Table~\ref{tSSE} gives the results for $\overline{SSe}$ when $|\gamma_{\rm oo}$ is artificially forced to zero. Here we see the fit is considerably improved by including the Coulomb correlation parameter in the fit. 

\subsubsection{Independence of fit results on tip height}
\label{sHeightQPWF}
Due to the very small lever arm $\alpha \sim10^{-2}$, the data is acquired very near flat-band\cite{Mol:2013dj}, and as such we do not observe the so-called ``ionization parabolas'' (observed elsewhere and associated with tip-induced band bending\cite{Teichmann:2008bh,Teichmann:2011gs}) for single B:Si acceptors\cite{Mol:2015im}, or for B:Si acceptor pairs (Fig. 1b, main text). 

\begin{figure}
\includegraphics{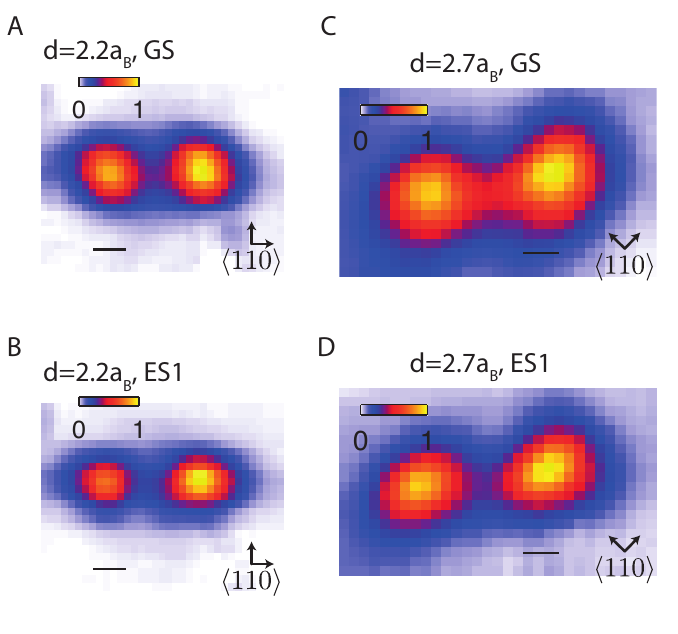}
\begin{flushleft}
\caption{A. Spatial measurement of ground state for $d=2.2a_{\rm B}$.  B.  Same as A, but for first excited state.  C.  Spatial measurement of ground state for $d=2.7a_{\rm B}$.  D.  Same as C, but for first excited state.  $\left\langle110\right\rangle$ crystal directions directions are as indicated. Scale bar: 1 nm}
\label{fDoubleBoron}
\end{flushleft}
\end{figure}

\begin{figure*}
\includegraphics{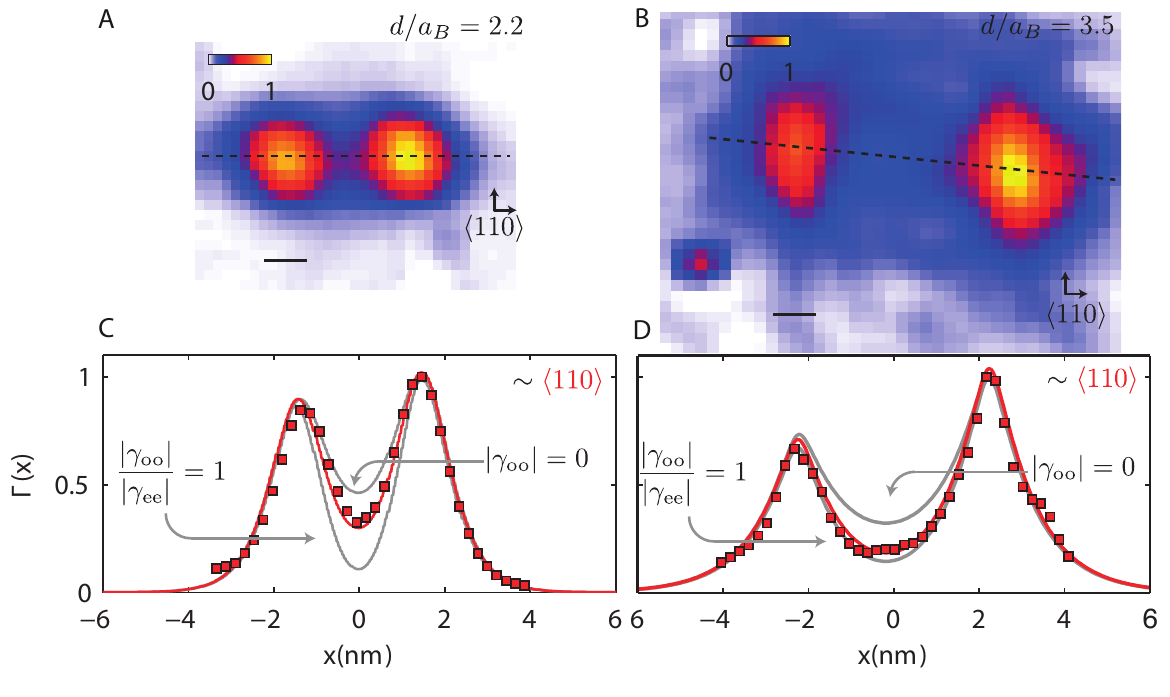}
\caption{A. Experimentally measured, normalized tunneling probability $\Gamma$ to tip, for $d=2.2a_{\rm B}$ GS.  Arrows denote $110$ crystal directions.  B. Same as (A), for $d=3.5a_{\rm B}$. C. Normalized experimental line profile (coloured squares) of $\Gamma(x)$ for $d=2.2a_{\rm B}$ and least-squares fit (coloured line) to correlated singlet model.  Upper and lower grey lines are line profiles of densities $|\phi_e(\mathbf{r})|^2$ and $|\phi_o(\mathbf{r})|^2$ respectively, obtained from least square fits.  D.  Same as (C), for $d=3.5a_{\rm B}$.   Scale bar: 1 nm}
\label{fCorr}
\end{figure*}

To provide an independent check that the tip does not influence the results of the QPWF measurement, we present measurements on the $d/a_{\rm B}=2.2$ and $d/a_{\rm B}=3.5$ pairs from the main text at a tip height $+60$ pm above the position where the data was taken in the main text.  The normalized tunneling probability is shown in Fig.~\ref{fCorr}A for $d/a_{\rm B}=2.2$ and in Fig.~\ref{fCorr}B for $d/a_{\rm B}=3.5$.  Line profiles of the normalized tunneling probability (squares, Fig.~\ref{fCorr}C and Fig.~\ref{fCorr}D), are in excellent agreement with the same QPWF model (solid red line, Fig.~\ref{fCorr}C and Fig.~\ref{fCorr}D) used in the main text.  Reference curves are shown for the totally uncorrelated state ($\gamma_{\rm oo}=0$) and maximally correlated state ($\gamma_{\rm oo}/\gamma_{\rm ee}=1$).  

The extracted results for the correlations are $|\gamma_{\rm oo}|^2=0.15\pm0.06$ for $d/a_{\rm B}=2.2$ and $|\gamma_{\rm oo}|^2=0.40\pm0.09$ for $d/a_{\rm B}=3.5$.  These results for $+60$ pm tip heights are within the experimental errors of the results for the data in the main text, which are $|\gamma_{\rm oo}|^2=0.12\pm0.06$ for $d/a_{\rm B}=2.2$ and $|\gamma_{\rm oo}|^2=0.39\pm0.08$ for $d/a_{\rm B}=3.5$.  

\subsubsection{Large distance limit of fitting model}
For any given experimental signal-to-noise ratio, there always exists a large enough $d/a_{\rm B}$ where it is impossible to distinguish the existence of correlations in the ground state QPWF.  This follows from the fact that our model determines correlations based on the contrast between orbitals $|\phi_{\rm e}(\mathbf{r})|^2$ and $|\phi_{\rm o}(\mathbf{r})|^2$.  However, since $|\phi_{\rm e}(\mathbf{r})|^2-|\phi_{\rm o}(\mathbf{r})|^2\sim 2\phi_A(\mathbf{r})\phi_B(\mathbf{r})$, where $\phi_{A/B}(\mathbf{r})$ are atomic orbitals at sites $A$ and $B$, and the product $2\phi_A(\mathbf{r})\phi_B(\mathbf{r})$ is exponentially suppressed with increasing $d/a_{\rm B}$.  The results of Table \ref{tSSE} show that the Coulomb correlations embodied by $|\gamma_{\rm oo}|$ can be reliably detected, for the signal-to-noise ratio of our experiment.

\subsubsection{Spatial measurements of excited states}
We present the ground and excited states of the acceptor pairs for $d/a_{\rm B}=2.2$ and $2.7$ in Fig.~\ref{fDoubleBoron}.  Compared with the ground states for $d/a_{\rm B}=2.2$ and $d/a_{\rm B}=2.7$ in Fig.~\ref{fDoubleBoron}A and Fig.~\ref{fDoubleBoron}C respectively, the excited states for $d=2.2a_{\rm B}$ (Fig.~\ref{fDoubleBoron}B) and $d=2.7a_{\rm B}$ (Fig.~\ref{fDoubleBoron}D) are slightly spatially elongated along the $110$ direction.  This is consistent with the expected QPWF density in the interpretation of Fig.~\ref{fig5} (main text), that the excited state orbitals have one hole occupying a ``1/2'' pseudospin state and the other occupying a ``3/2'' pseudospin state.  These states display slightly different spatial anisotropies (Fig.~\ref{fSingleBoron}C and Fig.~\ref{fSingleBoron}D).  

While we have measured the spatial tunneling probabilities for the excited states, measuring excited-state quasi-particle wavefunctions requires slow tunnel-in from the tip into the dopant system (not slow tunnel-out to the tip).  Slow tunnel-in from the tip is required to make the current limiting slow tunnel rates additive for excited states (rather than competitive with the ground state as in the case herein), in the presence of Coulomb blockade.  Some details are given in reference \onlinecite{Voisin:2015gl}.

\subsection{Tip-height dependent spectra}
\label{sTipHeightSpectra}

\begin{figure}
\includegraphics{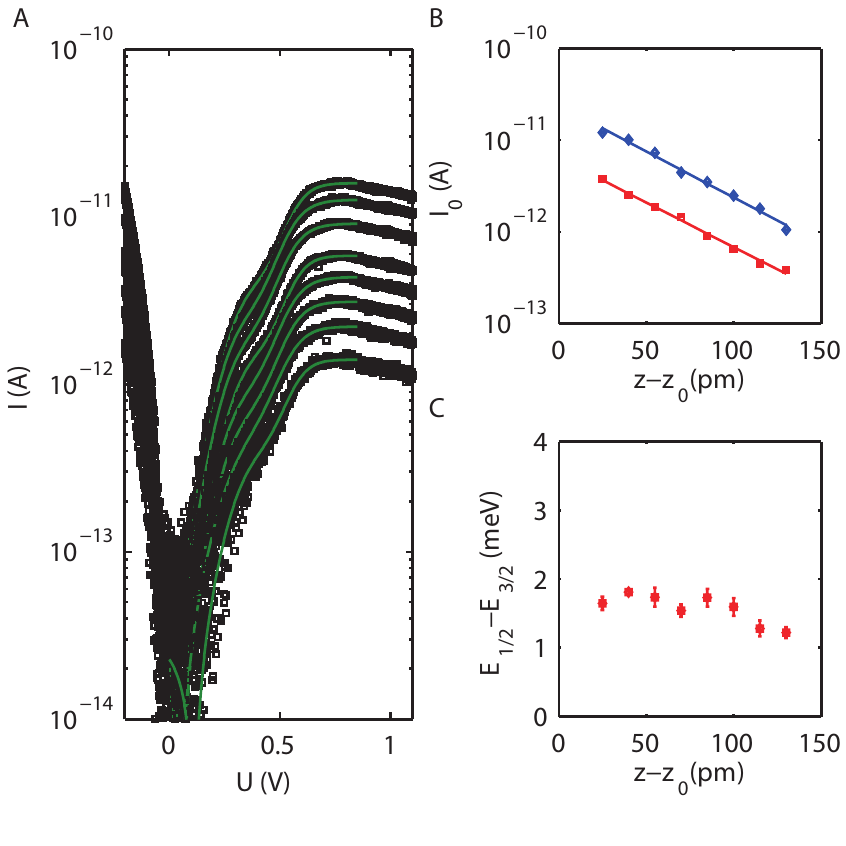}
\begin{flushleft}
\caption{A. Tip height dependence of current $I$ vs sample bias $U$ for single acceptor.  Two distinct thermally broadened steps corresponding to $\pm3/2$ states and $\pm1/2$ states are observed\cite{Mol:2015im}. B. Peak current $I_0$ reduces exponentially as a function of tip height. C. Energy splitting to the $\pm1/2$ excited state }
\label{fIVZ}
\end{flushleft}
\end{figure}

In this section we discuss our confirmation of our model for the thermally broadened sequential tunneling in Fig.~\ref{fdIdUfit}E, by detailed investigation of the tip-height dependence of the current, in agreement with our results on Arsenic donors in Si\cite{Salfi:2014kaa,Voisin:2015gl}.  This is shown in Fig.~\ref{fIVZ}A for tip heights that sequentially increase by $15$ pm per curve, above a single B:Si acceptor presenting the unique topographical signature discussed elsewhere\cite{Mol:2013dj,Mol:2015im}.  The data shows broadened current steps centred at $U\approx0.25$ V and $U\approx 0.5$ V that have an exponential dependence on the tip height, and which fit very well the lineshape of purely thermally broadened current resonance of a dopant\cite{Mol:2013dj,vanderHeijden:2014fp,Salfi:2014kaa,Voisin:2015gl,Mol:2015im}.  The peaks shown here correspond to the $\pm 3/2$ ground state and $\pm 1/2$ first excited state discussed in Section \ref{sSpatialModel} and reference \onlinecite{Mol:2015im}.     

Qualitatively, the exponential dependence on tip height verifies the assumption that $\Gamma_{\rm in} \gg \Gamma_{\rm out}$, as expected for the large vacuum tunneling barrier, and as required to obtain the results in Fig.~\ref{fig3} of the main text.  We performed least-squares fitting to extract the height $I_0$ of the thermally broadened current steps.  These $I_0$ are plotted as a function of tip height $z-z_0$ in Fig.~\ref{fIVZ}b for the $\pm3/2$ ground state (red curve) and the $\pm1/2$ excited state (blue curve).  The data are in excellent agreement with a z-dependence $\exp(-2\kappa(z-z_0))$ where $\kappa=(1.15\pm0.05)\times10^{10}$ m$^{-1}$.  This value is in good agreement with the expected value for $\kappa=\sqrt{2m_0\Phi_B}/\hbar$ for the  $\approx 5$ eV barrier expected for shallow valence band acceptor in silicon.  Moreover, we find that the lever arm (not shown) and energy splitting (Fig.~\ref{fIVZ}C) between the ground and excited state does is essentially independent of tip height.  This provides further verification of the sequential tunneling model, and that the tip does not strongly influence the acceptor-bound states.

\subsection{Hubbard model vs. Molecular orbitals}
\label{sHubbard}
The results for the two-site Hubbard model and its relationship to the molecular orbital model in Fig.~\ref{fig2}B of the main text are straightforward to derive. The coefficients $\gamma_{\rm c}$ and $\gamma_{\rm i}$ of the singlet ground state $\left|\Psi_S\right\rangle=\gamma_{\rm c}(\left|\uparrow;\downarrow\right\rangle-\left|\downarrow;\uparrow\right\rangle)+\gamma_{\rm i}(\left|\uparrow\downarrow;\right\rangle+\left|;\uparrow\downarrow\right\rangle)$ (Fig.~\ref{fig2}b, main text)  are readily found by direct diagonalization to be $\gamma_{c}=a/\sqrt{2+a^2}$ and $\gamma_{i}=1/\sqrt{2+2a^2}$ where $a=(-U/4t+\sqrt{1+(U/4t)^2})^{-1/2}$.

The dependence of $\gamma_{ee}$ and $\gamma_{\rm oo}$ on $U/t$ in Fig.~\ref{fig2}B can be found by a simple transformation.  The orthonormal localized states of the Hubbard dimer model created by operators $c_{A\sigma}^\dagger$ and $c_{B\sigma}^\dagger$ are 
\begin{align}
&c^\dagger_{A\sigma}=w^{-1/2}(c^\dagger_{a\sigma}-gc^\dagger_{b\sigma}) \textrm{ and}\\
&c^\dagger_{B\sigma}=w^{-1/2}(c^\dagger_{b\sigma}-gc^\dagger_{a\sigma})
\end{align}
respectively\cite{Schliemann:2001iz}.  Here $c^\dagger_{a\sigma}$ and $c^\dagger_{b\sigma}$ create (non-orthogonal) atomic orbitals with spin $\sigma$, $w=1-2Sg+g^2$, $g=(1-\sqrt{1-S^2})/S$, and $S=\left\langle a|b\right\rangle$ is the overlap between the atomic orbitals. 

Re-writing the even and odd eigenstates of the single-particle potential and equation the two singlet ground states we find $\gamma_{\rm ee}=\gamma_{\rm c}+\gamma_{\rm i}$ and $\gamma_{\rm oo}=\gamma_{\rm c}-\gamma_{\rm i}$.  These relationships are used to map the Hubbard model solution to the molecular orbital model in Fig.~\ref{fig2}B of the main text.  This mapping is independent of normalization constants $w$, $g$, and $S$. 

\subsection{Degree of entanglement}
\label{sEntanglement}

For fermions, entanglement is defined in terms of Slater decompositions\cite{Amico:2008en} rather than the Schmidt decompositions of distinguishable particles, and can be expressed independent of basis by a degree of entanglement.  Here we follow derivation in reference [\onlinecite{He:2005in}] for entanglement to describe correlations in a two-site Fermi-Hubbard system.  For two particles that obey Fermi statistics,
\begin{equation}
|\Psi\rangle=\sum_{a,b}\omega_{a,b}|a\rangle_1 |b\rangle_2
\end{equation}
is characterized by an anti-symmetric matrix $\omega_{a,b}$, that is, a Slater decomposition.  Transformed into a block diagonalized form $\textrm{diag}[Z_0,Z_1,...,Z_N]$ through a unitary rotation of the single particle states, where 
\begin{equation}
Z_i= \left( \begin{array}{cc}
0 & z_i\\
-z_i & 0\end{array} \right),
\end{equation}
the number of nonzero $z_i$ is called the Slater rank, and if the Slater rank is 1, the quantum correlation of the state is zero\cite{Schliemann:2001ea}.  It has been shown\cite{He:2005in} that the $z_i^2$ are the eigenvalues of the basis independent quantity $\omega^\dagger\omega$, and based on this observation, the degree of entanglement $\mathcal{S}$ of the two particles was defined as
\begin{equation}
\mathcal{S}=-\sum_i z_i^2 \log_2 z_i^2.
\end{equation}

For a superposition of ``even/even'' and ``odd/odd'' singlets $|S\rangle=(\gamma_{\rm ee} c_{\rm e,\uparrow}^\dagger c_{\rm e,\downarrow}^\dagger-\gamma_{\rm oo} c_{\rm o,\uparrow}^\dagger c_{\rm o,\downarrow}^\dagger)|0\rangle$, 
\begin{equation}
\omega= \left( \begin{array}{cccc}
0 & \gamma_{\rm ee} & 0 & 0\\
-\gamma_{\rm ee} & 0 & 0 & 0\\
0 & 0 & 0 & -\gamma_{\rm oo}\\
0 & 0 & \gamma_{\rm oo} & 0\end{array} \right).
\end{equation}
It directly follows that the non-zero eigenvalues $z_i^2$ of $\omega^\dagger\omega$ are $|\gamma_{\rm ee}|^2$ and $|\gamma_{\rm oo}|^2$.  Then, the degree of entanglement for the state $\left|S\right\rangle$ is
\begin{equation}
\mathcal{S} = -|\gamma_{\rm ee}|^2 \log_2|\gamma_{\rm ee}|^2 - |\gamma_{\rm oo}|^2 \log_2|\gamma_{\rm oo}|^2.
\end{equation}

%\bibliography{exchangedopants_ncomm}

\end{document}